\documentclass[iop]{emulateapj}
\begin{document}

\newcommand{\kms}{km\ s$^{-1}$}
\newcommand{\kmsmpc}{km\ s$^{-1}$\ Mpc$^{-1}$}
\newcommand{\ergscm}{erg\ s$^{-1}$ cm$^{-2}$} 
\newcommand{\OII}{[\ion{O}{2}]}
\newcommand{\Deg}{^{\circ}}

\title{The X-ray--Optical Relations for Nine Clusters at $z$ = 0.7 - 1.1 from the ORELSE Survey}

\author{
  N. Rumbaugh\altaffilmark{1}, D. D. Kocevski\altaffilmark{2}, R. R. Gal\altaffilmark{3}, B. C. Lemaux\altaffilmark{4}, L. M. Lubin\altaffilmark{1}, C. D. Fassnacht\altaffilmark{1}, G. K. Squires\altaffilmark{5}
}

\email{Electronic address: narumbaugh@ucdavis.edu}

\altaffiltext{1}{Department of Physics, University of California,
  Davis, 1 Shields Avenue, Davis CA 95616, USA}
\altaffiltext{2}{Department of Physics and Astronomy, University of Kentucky,
  Lexington, KY 40506-0055, USA}
\altaffiltext{3}{University of Hawai'i,
    Institute for Astronomy, 2680 Woodlawn Drive, HI 96822, USA}
\altaffiltext{4}{Aix Marseille Universit\'{e}, 
    CNRS, Laboratoire d'Astrophysique de Marseille, UMR 7326, 13388, Marseille, France}
\altaffiltext{5}{California Institute of Technology, M/S
    220-6, 1200 E. California Blvd., Pasadena, CA 91125, USA}

\begin{abstract}

We use {\it Chandra} observations of nine optically and X-ray selected clusters in five different structures at $z\sim0.7-1.1$ from the Observations of Redshift Evolution in Large-Scale Environments (ORELSE) survey to study diffuse X-ray emission from galaxy clusters. X-ray gas temperatures and bolometric rest-frame luminosities are measured for each cluster in the sample. We present new redshift measurements, derived from data obtained using the Deep Imaging Multi-Object Spectrograph on the Keck 10-m telescope, for two clusters in the RX J0910 supercluster at $z\sim1.1$, from which velocity dispersions are measured. Dispersions for all clusters are combined with X-ray luminosities and gas temperatures to evaluate how the cluster properties compare to low-redshift scaling relations. We also measure the degree of substructure in each cluster by examining the velocity histograms, performing Dressler-Shectman tests, and computing the offsets between the X-ray emission center and optically-derived centroids. We find that only two clusters show clear indications of being unrelaxed, based on their scaling relations and other dynamical state diagnostics. Using our sample, we evaluate the redshift evolution of the $L_x$-$T$ relation and investigate the implications of our results for precision cosmology surveys. 

\end{abstract}

\section{Introduction}

Galaxy clusters can be used to constrain cosmological parameters such as $\sigma_8$ or the dark energy equation of state by, for example, measuring their abundances or mass function \citep{vik09,rozo10,allen11}. As the largest virialized structures in the universe, clusters trace the large-scale structure in the universe, and their distribution is a test for models of structure formation and evolution. In order to use galaxy clusters as tools, we require reliable estimates of their masses. Since mass is not an observable quantity, other measureables must be used as proxies. Scaling relations between the fundamental physical property - total mass - and observables, such as the X-ray luminosity or temperature of the intracluster medium (ICM), or the composite parameter $Y_x$, are therefore essential \citep[e.g,][]{krav06,vik09b}.

Assumptions of self-similarity can simplify the use of scaling relations between cluster properties. However, studies at low redshift find deviations from self-similarity among virialized clusters, suggesting the influence of non-gravitational processes, such as active galactic nuclei (AGN) feedback, on the formation and evolution of clusters \citep{mark98,XW00,mc11}. We would expect non-gravitational heating to vary with time, which would affect the redshift evolution of the scaling relations. A number of studies have sought to evaluate this evolution with observations at higher redshifts. Results have been mixed, with some studies finding consistency with self-similar evolution, while others find deviations from it \citep[see e.g.,][and references therein]{maug11,rei11}. \citet{rei11}, using an extensive $z\lesssim1.5$ sample compiled from the literature, find that the $L_x$-$T$ relation scales as $E\left(z\right)^{\alpha}$, with $\alpha = -0.23^{+0.12}_{-0.62}$, in contrast to the $\alpha = 1$\ prediction from self-similarity. As with the deviations found in local scaling relations, this result would indicate a significant heating contribution from non-gravitational sources, although different processes may be important at higher redshifts. 

Scaling relations between cluster properties are often calibrated using local virialized clusters \citep[e.g.,][]{XW00,arnaud05}. These relations are not expected to provide an accurate characterization of unrelaxed clusters that are still undergoing gravitational heating \citep{cast11}. Identifying unrelaxed clusters is therefore useful when deriving cluster properties based on scaling relations. While many studies use measures of morphology to identify unrelaxed clusters \citep[e.g.,][]{pratt09,vik09b,maug11}, the efficacy of such tests varies and is in need of further study. As an illustration of the uncertain techniques of identifying unrelaxed clusters, \citet{lopes06}, using a sample of 618 clusters with $0.05 \le z \le 0.40$, find that the fraction of their sample that contains substructure varies from 13\% to 45\% for four different optical tests. 

In this paper, we use a sample of nine clusters in the range $0.7<z<1.1$, initially selected using both X-ray and optical techniques, in five large-scale structures observed as part of the Observations of Redshift Evolution in Large-Scale Environments (ORELSE) survey \citep{lub09}. We use {\it Chandra} observations to search for diffuse emission around clusters in our sample. We determine X-ray gas temperatures and luminosities of the ICM for these clusters and study scaling relations between these properties and the velocity dispersions derived from confirmed cluster galaxies. Using several diagnostic tests, we determine the dynamical states of the members of our sample. For our cosmological model, we assume $\Omega_m = 0.3$, $\Omega_\Lambda = 0.7$, and $h_{70} =H_0/70\ $\kmsmpc.

In Section \ref{sec:sample}, we provide an overview of the structures comprising our sample. In Section \ref{sec:red}, we describe the optical and X-ray observations. In Section \ref{sec:clusprop}, we discuss our measurements of cluster properties. We first cover measurements of cluster centroids followed by the velocity histograms and dispersions and the Dressler-Shectman tests for substructure. In addition, we outline our search for diffuse X-ray emission within the fields and analyze detected emission. In Section \ref{sec:scalrel}, we discuss scaling relations between the different cluster properties. In Section \ref{sec:anal}, we analyze the results of previous measurements and diagnostic tests and explore their implications for our sample and for other cluster surveys. In Section \ref{sec:con}, we summarize and discuss our results. 

\section{The Sample}
\label{sec:sample}

We analyze {\it Chandra} observations of a total of five large-scale structures: RX J1821.6+6827, RX J1757.3+6631, the RX J0910 supercluster, the Cl 1324 supercluster, and the Cl 1604 supercluster. Of these, {\it Chandra} data have only been used to study diffuse emission for the Cl 1604 supercluster \citep{koc09a} and RX J0910+5422 \citep{stan02}, which is part of the RX J0910 structure. Here we review the characteristics of these five structures.

\subsection{RX J1821.6+6827}

RX J1821.6+6827, hereafter RX J1821, is an X-ray selected cluster at $z=0.82$. RX J1821 was discovered as part of the ROSAT \citep{ROSAT} North Ecliptic Pole (NEP) survey and was the highest redshift cluster therein \citep{gioia03,hen06}. \citet{gioia04} studied diffuse emission from the cluster using XMM-Newton data \citep{XMM}. They measured a bolometric X-ray luminosity of $1.17^{+0.13}_{-0.18} \times 10^{45}\ h_{70}^{-2}$\ ${\rm erg}\ s^{-1}\ $and a temperature of $4.7^{+1.2}_{-0.7}\ $keV and found the emission to be slightly elongated. The temperature measured by \citet{gioia04} is consistent within the errors with our results, although the bolometric X-ray luminosity is a factor of two smaller (see Table \ref{specanntab}), most likely due to AGN contamination in the XMM data. Analysis of the redshift histogram has found RX J1821 to be dominated by a single, large structure, with a small kinematically associated group detected to the south \citep{lub09}. \citet{rum12} reported a velocity dispersion within 1 $h_{70}^{-1}\ $Mpc for the cluster of $910\pm80$ \kms\ using 42 galaxies.

We refer the reader to \citet{lub09}, \citet{lem10}, and \citet{rum12} for more information on the structure and observations.  

\subsection{RX J1757.3+6631}

RX J1757.3+6631, hereafter RX J1757, is an X-ray selected cluster at a redshift of $z=0.69$\ discovered in the ROSAT NEP survey \citep{gioia03}. While no gas temperature has previously been published for the structure, \citet{gioia03} measured an X-ray luminosity of 8.6$\times 10^{43}\ h^{-2}_{70}\ $erg s$^{-1}$ in the 0.5-2.0 keV band\footnote{This ROSAT measurement may have a large error due to the presence of two X-ray bright AGN close to the cluster core \citep{gioia03}.}. The structure is dominated by a single, large cluster, with a measured velocity dispersion of $650\pm120$ \kms\ within 1 $h_{70}^{-1}\ $Mpc using 21 galaxies \citep{rum12}. 

We refer the reader to \citet{rum12} for more information on this structure and the observations. 

\subsection{The RX J0910 Supercluster}
\label{sec:0910}

The first discovered cluster in this structure, RX J0910+5422, was selected from the ROSAT Deep Cluster Survey \citep{ros95} at $z=1.1$. Observations by \citet{stan02} show a red galaxy overdensity whose peak is consistent with that of the diffuse X-ray emission. Using {\it Chandra} data, they measured a bolometric X-ray luminosity of $2.48^{+0.30}_{-0.26} \times h_{70}^{-2}\ {\rm erg}\ s^{-1}\ $and a temperature of $7.2^{+2.2}_{-1.4}\ $keV, consistent within $\sim 2\sigma$\ with our measurements. \citet{stan02} noted elongation in both the distribution of cluster members and the diffuse emission, suggesting the cluster is still in the process of forming. \citet{mei06} studied the cluster using color magnitude diagrams (CMDs) constructed using Hubble Space Telescope (HST) data. They found that the S0 population was significantly bluer than the elliptical galaxies, which could also support the conclusion that the cluster is forming. An extensive spectroscopic campaign by \citet{tan08} confirmed RX J0910+5422 as a bound cluster at $z=1.101\pm0.002$. Wide-field optical and X-ray imaging revealed another cluster, RX J0910+5419, within close proximity, $\sim 6'$, on the sky. \citet{tan08} also found evidence of filaments and other potential clusters and groups, suggesting the presence of large-scale structure. Hereafter, we will refer to the structure as a whole as RX J0910. 

Because of the suggestions of large-scale structure present in RX J0910, it was chosen as part of the ORELSE survey. In this paper, we present new spectroscopic results for the structure, coupled with those from \citet{tan08}.

\subsection{The Cl 1324 Supercluster}

The Cl 1324 supercluster at $z\approx0.76$\ spans about 25$\ h_{70}^{-1}$ Mpc on the plane of the sky and $110\ h_{70}^{-1}$ Mpc along the line of sight. It was first discovered as two overdensities in the survey of \citet{gunn86}. These overdensities correspond to the two largest clusters in the structure, Cl 1324+3011 at redshift $z=0.76$, and Cl 1324+3059 at redshift $z=0.69$. Because of the proximity of the overdensities, the structure was investigated as part of the ORELSE survey \citep[see Gal et al. 2013, in preparation; ][]{rum12}. Wide-field imaging has detected ten clusters and groups through red-galaxy overdensities, although only four have been spectroscopically confirmed. 

Cl 1324+3011 was previously studied by \citet{lub02,lub04}. They measured a velocity dispersion of $1016^{+126}_{-93}$ \kms, using 47 galaxies, and a temperature of $2.88_{-0.49}^{+0.71}\ $keV using XMM-Newton, consistent with our {\it Chandra} measurement (see Table \ref{specanntab}). These results imply the cluster is not well relaxed, as it lies off the $\sigma_v -T$ curve for virialized clusters. \citet{rum12} present new velocity dispersion measurements for four clusters in the structure. Three of these clusters are studied here, with updated velocity dispersion measurements, listed in Table \ref{specanntab} (see Section \ref{sec:veldisp} for more details).

\subsection{The Cl 1604 Supercluster}

The Cl 1604 supercluster at $z\approx 0.9$\ is one of the largest structures studied at high redshifts. The structure contains 10 detected clusters and groups and spans $13\ h_{70}^{-1}\ $Mpc along the line of sight and $100\ h_{70}^{-1}\ $Mpc in the plane of the sky \citep{lubin00,GalLub04,gal08,lemaux09}. The structure, like Cl 1324, was first discovered as two clusters, Cl 1604+4304 and Cl 1604+4321, in the survey of \citet{oke98}. Through wide field imaging, 10 distinct red-galaxy overdensities have been detected in the supercluster \citep{lubin00,GalLub04,gal08}. Three of the overdensities are clusters with velocity dispersions in excess of 500 \kms, while five others are poor clusters or groups with dispersions in the range 300-500 \kms\ \citep{post98,post01,gal05,gal08}. 

Diffuse X-ray emission from the clusters and groups in the supercluster has been studied previously by \citet{koc09a}. Emission was detected for Cl 1604+4304 and Cl 1604+4314, hereafter Cl 1604A and Cl 1604B, with measured bolometric X-ray luminosities of $15.76 \pm 1.48\ $ and $11.64\pm  1.49 \times 10^{43}\ h_{70}^{-2}\ $erg s$^{-1}\ $ and X-ray temperatures of $3.50^{+1.82}_{-1.08}\ $ and $1.64^{+0.65}_{-0.45}\ $keV, respectively \citep{koc09a}. No diffuse emission was detected from any other group or cluster in the supercluster, which places an upper limit on their bolometric X-ray luminosities of approximately $7.4\times 10^{43}\ h_{70}^{-2}\ $erg s$^{-1}\ $\citep{koc09a}. 

We refer the reader to \citet{koc09a}, \citet{koc09b}, \citet{gal08}, and \citet{lem11} for more details on the structure and the optical observations. 

\section{Observations and Reduction}
\label{sec:red}

The data that we analyze in this paper include optical imaging and spectroscopy and {\it Chandra} X-ray observations. The observations, excluding those of the RX J0910 supercluster, are presented in more detail in \citet{rum12}.

\subsection{Optical Imaging}
\label{sec:optim}

Ground-based optical imaging was carried out for all fields using the Large-Format Camera \citep[LFC;][]{simcoe00}\ on the Palomar 5m Telescope in the $r'$, $i'$, and $z'$ bands. The Cl 1604 field has additional data in the F606W and F814W bands from the Advanced Camera for Surveys (ACS) on the HST.

We used our optical imaging to determine the red sequences in each field. Red sequence fits were performed for each structure and were calculated using a linear fitting and $\sigma$-clipping technique as in \citet{rum12}. First, a fit to a linear model was carried out on member galaxies within a chosen magnitude and color range using a $\chi^2$ minimization \citep{gladders98,stott09}. The fit was initialized with a color range chosen ``by eye'' to conform to the apparent width of the red sequence of the structure. After an initial fit, colors were normalized to remove the slope. The color distribution was then fit to a single Gaussian using iterative $3\sigma$ clipping. At the conclusion of the algorithm, the boundaries of the red sequence were defined by a 3$\sigma\ $offset from the center, except for Cl 1604 and Cl 1324. The color dispersion for these structures was inflated due to their large redshift extents, and 2$\sigma$ offsets were used to achieve reasonable boundaries.

CMDs for the structures in our sample, except for RX J0910, can be found in \citet{rum12}. 

\subsection{Optical Spectroscopy}

Our photometric catalogs are complemented by an unprecedented amount of spectroscopic data for large-scale structures at high redshifts. These data were taken using the Deep Imaging Multi-Object Spectrograph \citep[DEIMOS;][]{faber03}\ on the Keck II 10m telescope, as described in \citet{rum12}, and reduced using the DEIMOS Data Reduction pipeline, {\it spec2d} \citep{coop12,new12}. RX J0910 has additional data, as described in the next section, and Cl 1604, Cl 1324, and RX J1821  have some coverage by the Low-Resolution Imaging Spectrograph \citep[LRIS;][]{oke95}, as well. Also, since the presentation of the data in \citet{rum12}, additional spectroscopy was obtained for RX J1821 and RX J1757, with slitmasks designed to preferentially target X-ray sources. We present updated velocity dispersions including these new data. In total, our observations have provided 1849 high-quality extragalactic spectra (Q=3,4; See \citealt{gal08} and \citealt{rum12} for more details) for Cl 1604, 1156 for Cl 1324, 539 for RX J1757, and 422 for RX J1821. From these data, we now have a total of 531 confirmed members for Cl 1604, 393 for Cl 1324, 54 for RX J1757, and 103 for RX J1821.

For more details of the spectroscopic observations, excluding those of RX J0910, we refer the reader to \citet{rum12}. 

\subsubsection{RX J0910 Spectroscopy}

Coverage of the RX J0910 supercluster includes DEIMOS data, LRIS data, and data from the Faint Object Camera and Spectrograph \citep[FOCAS;][]{kash02}\ on Subaru. Details of the latter two datasets are given in \citet{mei06}\ and \citet{tan08}, respectively. As part of ORELSE, we have obtained DEIMOS data on the RX J0910 field. We used the 1200 line mm$^{-1}$ grating, tilted to a central wavelength of 8000-8100 \AA, and 1$\arcsec$ slits. Exposure times were in the range 9000-10800s. The DEIMOS observations yielded 459 high-quality extragalactic spectra, while the previous LRIS and FOCAS observations contain an additional 131. 
\begin{deluxetable}{lcccc}
\tablecaption{{\it Chandra} Observation Characteristics}
\tablehead{
\colhead{\footnotesize{Structure}}
 & \colhead{\footnotesize{Obs.}}
 & \colhead{\footnotesize{Pointing}}
 & \colhead{\footnotesize{Pointing}}
 & \colhead{\footnotesize{Exposure}} \\
 \colhead{\footnotesize{}} 
& \colhead{\footnotesize{ID}}
 & \colhead{\footnotesize{R.A.(J2000)}}
 & \colhead{\footnotesize{Dec.(J2000)}}
& \colhead{\footnotesize{Time (ks)}}
}
\startdata
Cl 1324 & \ 9403 & 13\ 24\ 48.9 & 30\ 51\ 49 & \ 26.9\\
Cl 1324 & \ 9840 & 13\ 24\ 48.9 & 30\ 51\ 49 & \ 21.5\\
Cl 1324 & \ 9404 & 13\ 24\ 42.0 & 30\ 16\ 46 & \ 30.4\\
Cl 1324 & \ 9836 & 13\ 24\ 42.0 & 30\ 16\ 46 & \ 20.0\\
Cl 1604 & \ 6932 & 16\ 04\ 19.7 & 43\ 10\ 14 & \ 49.5\\
Cl 1604 & \ 6933 & 16\ 04\ 10.5 & 43\ 22\ 33 & \ 26.7\\
Cl 1604 & \ 7343 & 16\ 04\ 10.5 & 43\ 22\ 33 & \ 19.4\\
RX J1821 & 10444 & 18\ 21\ 13.4 & 68\ 27\ 48 & \ 22.2\\
RX J1821 & 10924 & 18\ 21\ 13.4 & 68\ 27\ 48 & \ 27.3\\
RX J1757 & 10443 & 17\ 57\ 19.5 & 66\ 29\ 23 & \ 21.7\\
RX J1757 & 11999 & 17\ 57\ 19.5 & 66\ 29\ 23 & \ 24.7\\
RX J0910 & \ 2227  & 09\ 10\ 40.0 & 54\ 19\ 57 & 107.0\\
RX J0910 & \ 2452  & 09\ 10\ 40.0 & 54\ 19\ 57 & \ 66.2\\
\enddata
\label{ChanObsTab}
\end{deluxetable}

\begin{figure}
\epsscale{1.1}
\plotone{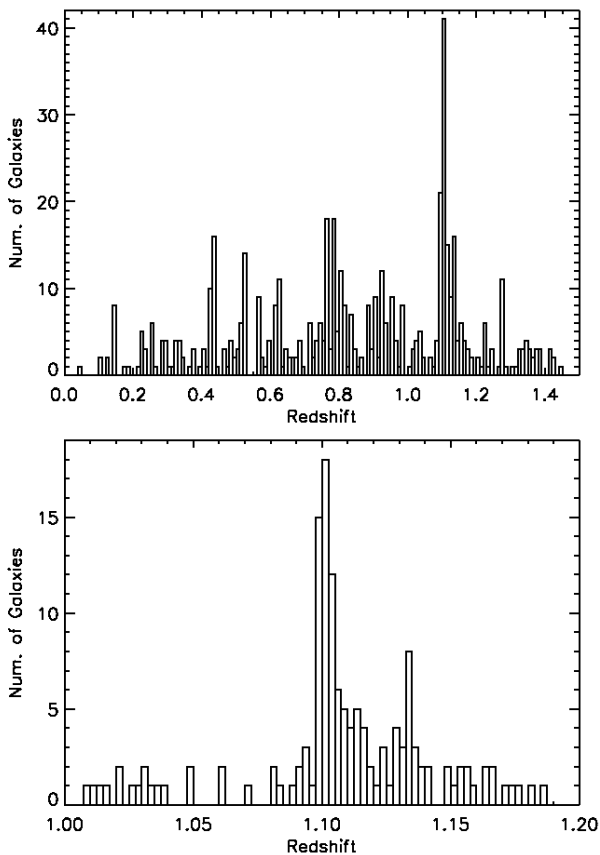}
\caption{\footnotesize{
Redshift histograms of RX J0910, including only reliable redshifts. In the top panel, we show all redshifts in the field below $z = 1.5$. In the bottom panel, only redshifts in the range $1.0 < z < 1.2$ are shown.
}}
\label{rsfig}
\end{figure}

Redshifts derived from all of the available spectroscopic observations for RX J0910 are shown in Figure \ref{rsfig}. In the top panel are all reliable redshifts with $z < 1.5$. A clear peak is visible at the redshift of the RX J0910 supercluster ($z \approx 1.1$). Several smaller peaks are visible as well at redshifts of $z\sim\ $0.4, 0.55, and 0.8. Upon examination, galaxies corresponding to each of these peaks appear to be uniformly spread across the field, implying they are mass sheets. In the lower panel, we plot a histogram of galaxies with $1.0 < z < 1.2$. The peak at $z \approx 1.1$ is dominated by the two clusters described in Section \ref{sec:0910} and appears approximately Gaussian. A Kolmogorov-Smirnov (K-S) Test does not reject an underlying normal distribution. There appears to be one other, smaller peak at $z \approx 1.13$. Upon examination, galaxies corresponding to this peak are uniformly spread across the field, suggesting that it does not represent a localized structure. 

We refer the reader to \citet{rum12} and references therein for the observations and discussion of the other fields. 

\subsection{{\it Chandra} Observations}
\label{sec:ChanObs}

All X-ray imaging of the clusters was conducted with the Advanced CCD Imaging Spectrometer (ACIS) of the {\it Chandra} X-ray Observatory, using the ACIS-I array. This array has a $16\farcm9 \times 16\farcm9$ field of view. Since RX J0910, RX J1821, and RX J1757 have angular sizes smaller than the array size, each was imaged with one pointing of the array. However, Cl 1604 and Cl 1324, with angular sizes in excess of 20$\arcmin$, were observed with two pointings each. For Cl 1604, the two pointings are meant to cover as much of the structure as possible, and there is a small overlap of $\sim30\ $arcminutes$^2$. For Cl 1324, the two pointings are centered near the two largest clusters, Cl 1324+3011 and Cl 1324+3059. There is an approximately 13$\arcmin$ gap between the two pointings. Characteristics of the observations are listed in Table \ref{ChanObsTab}. Although exposure times for individual observations vary, each pointing has an effective exposure time of approximately 50ks, except for RX J0910 with a total exposure time of 175ks. While all {\it Chandra} observations from \citet{rum12} are described in this paper, observations of the Cl 0023 supergroup are not included in Table \ref{ChanObsTab} or the subsequent discussion because of the lack of detected diffuse emission in this structure. Cl 0023 was observed with ObsId 7914, which also had an approximate effective exposure time of 50 ks.

The reduction of the data was conducted using the Chandra Interactive Analysis of Observations 4.2 software \citep[CIAO;][]{frusc06}. Reduction was carried out separately in three bands: 0.5-2 keV (soft), 2-8 keV (hard), and 0.5-8 keV (full). Data were checked for flares and vignetting corrected. Point sources were located with the routine {\it wavdetect} and removed using the CIAO tool {\it dmfilth}. For more details of the X-ray data reduction, see \citet{rum12}.

\section{Cluster Properties}
\label{sec:clusprop}

In this section, we examine the properties of the clusters in our sample. We determine the cluster centers with a variety of techniques and study the diffuse X-ray emission from each cluster. 

\subsection{Optical Cluster Centroid Measurements}
\label{sec:optgalcen}

\begin{deluxetable*}{lcccccccccc}
\tablecaption{Optical Galaxy Centroid Measurements}
\tablehead{
\colhead{\footnotesize{Cluster}}
& \colhead{\footnotesize{BCG}}
& \colhead{\footnotesize{BCG}}
& \colhead{\footnotesize{BCG}}
& \colhead{\footnotesize{Red Gal.}}
& \colhead{\footnotesize{Red Gal.}}
& \colhead{\footnotesize{Red Gal.}}
& \colhead{\footnotesize{Mean Gal.}}
& \colhead{\footnotesize{Mean Gal.}}
& \colhead{\footnotesize{Mean}} \\
\colhead{\footnotesize{}}
& \colhead{\footnotesize{RA}}
& \colhead{\footnotesize{Dec.}}
& \colhead{\footnotesize{Offset From}}
& \colhead{\footnotesize{Peak RA}}
& \colhead{\footnotesize{Peak Dec.}}
& \colhead{\footnotesize{Peak}}
& \colhead{\footnotesize{RA}}
& \colhead{\footnotesize{Dec.}}
& \colhead{\footnotesize{Gal. Pos.}}\\
\colhead{\footnotesize{}}
& \colhead{\footnotesize{(J2000)\tablenotemark{a}}}
& \colhead{\footnotesize{(J2000)\tablenotemark{a}}}
& \colhead{\footnotesize{X-ray Cen.\tablenotemark{b}}}
& \colhead{\footnotesize{(J2000)\tablenotemark{c}}}
& \colhead{\footnotesize{(J2000)\tablenotemark{c}}}
& \colhead{\footnotesize{Offset\tablenotemark{d}}}
& \colhead{\footnotesize{(J2000)\tablenotemark{e}}}
& \colhead{\footnotesize{(J2000)\tablenotemark{e}}}
& \colhead{\footnotesize{Offset\tablenotemark{f}}}
}
\startdata   
   Cl 1324+3011 &  13 24 48.8 & 30 11 39.3 & \phm{1}98 &  13 24 48.8 & 30 11 39.3 & \phm{1}98 &  13 24 48.7 & 30 11 53.8 & 205 \\
  Cl 1324+3013 &  13 24 20.9 & 30 12 43.5 & \phm{1}86 &  13 24 21.6 & 30 12 53.9 & 126 &  13 24 21.2 & 30 12 56.7 & \phm{1}89 \\
  Cl 1324+3059 &  13 24 47.6 & 30 58 48.4 & 177 &  13 24 49.8 & 30 58 25.4 & \phm{1}87 &  13 24 48.2 & 30 58 18.9 & 146 \\
      RX J1757 &  17 57 19.6 & 66 31 32.9 & \phm{1}30 &  17 57 18.8 & 66 31 37.5 & \phm{1}64 &  17 57 20.5 & 66 31 26.7 & \phm{1}52 \\
      RX J1821 &  18 21 33.1 & 68 27 56.3 & \phm{1}35 &  18 21 32.1 & 68 28 16.3 & 147 &  18 21 31.4 & 68 28 21.5 & 190 \\
      Cl 1604A &  16 04 25.0 & 43 04 52.3 & 163 &  16 04 21.5 & 43 04 34.1 & 177 &  16 04 22.4 & 43 04 56.5 & 167 \\
      Cl 1604B &  16 04 26.2 & 43 14 19.1 & \phm{1}33 &  16 04 23.7 & 43 14 07.8 & 257 &  16 04 25.6 & 43 14 19.7 & \phm{1}81 \\
 RX J0910+5419 &  09 10 08.6 & 54 18 59.8 & \phm{1}31 &  09 10 04.2 & 54 18 54.2 & 309 &  09 10 02.7 & 54 18 33.8 & 454 \\
 RX J0910+5422 &  09 10 45.9 & 54 22 07.6 & \phm{1}67 &  09 10 47.7 & 54 22 13.8 & 199 &  09 10 44.6 & 54 22 20.8 & 115
\enddata
\label{opcentable}
\tablenotetext{a}{\footnotesize{Positions on the sky of the cluster BCG.}}
\tablenotetext{b}{\footnotesize{Offset, measured in $h_{70}^{-1}$ kpc, between the position of the cluster BCG and the centroid determined from diffuse X-ray emission contours (see Table \ref{specanntab}). Note that this distance cannot be more than 250 $h^{-1}_{70}$ kpc, by definition (see Section \ref{sec:BCG}).}}
\tablenotetext{c}{\footnotesize{Positions on the sky of the red galaxy density peaks corresponding to each cluster (see Section \ref{sec:RGDP} for more details).}}
\tablenotetext{d}{\footnotesize{Offset, measured in $h_{70}^{-1}$ kpc, between the red galaxy density peaks and the centroid determined from diffuse X-ray emission contours.}}
\tablenotetext{e}{\footnotesize{Positions on the sky of the mean luminosity-weighted position of all galaxies used in velocity dispersion measurements for the respective cluster (see Section \ref{sec:LWMC} for more details). }}
\tablenotetext{f}{\footnotesize{Offset, measured in $h_{70}^{-1}$ kpc, between mean luminosity-weighted position of galaxies in each cluster and the centroid determined from diffuse X-ray emission contours.}}
\end{deluxetable*}

There are many alternate methods for determining the center position for a cluster. Here we explore a variety of techniques using our optical data, including the peak of the smoothed red galaxy density distribution, the position of the brightest cluster galaxy (BCG), and the galaxy luminosity-weighted centroid. 

These measurements are given in Table \ref{opcentable} and are described below. 

\subsubsection{Red Galaxy Density Peaks}
\label{sec:RGDP}

Since red galaxies are more likely to be found in clusters than in the field and in cluster cores than in their outskirts \citep[e.g.,][]{HH31,dress80,WGJ93}, they can be used to both locate and to centroid galaxy clusters. We measured the positions of red galaxy density peaks for each cluster. Optical catalogs were filtered based on $r'-i'$ colors and $i'$ ranges designed to select red sequence galaxies at the redshifts of the structures. These galaxies were used to construct an adaptive kernel surface density map using a two-stage process, as in \citet{gal05}. Galaxy density was initially estimated on a fixed grid, which was then used to initialize adaptive smoothing. SExtractor was used to detect and centroid the peaks in the smoothed galaxy density map. The locations of these peaks are listed in Table \ref{opcentable}.

\subsubsection{Brightest Cluster Galaxies}
\label{sec:BCG}

BCGs tend to be located at the centers of massive clusters, both on the sky and in velocity space \citep{QL82,JF84}. Therefore, they can be used as estimates of the location of each cluster center. Additionally, a BCG with a large peculiar velocity or with an offset from the center of the cluster gas can be an indicator of a recently disturbed or not yet fully formed cluster \citep{bird94,GB02}. To locate BCGs, we searched within 0.25 $h_{70}^{-1}\ $Mpc of the X-ray centers considering only galaxies on the red sequence and with magnitudes in ranges that were reasonable for red sequence galaxies at the appropriate redshift. In all cases, we confirmed that we did not miss any more luminous member galaxies bluer than the red sequence. We have complete spectroscopy for the galaxies chosen as BCGs in our sample. The positions of the BCGs are listed in Table \ref{opcentable}. 

Identification of the BCGs was straightforward in all but one case. The object chosen as the BCG in RX J0910+5419, while very close to the center of X-ray emission, has a large velocity relative to the mean redshift of the cluster ($\sim 1800\ $\kms). The next brightest galaxy has very similar $r'$ and $i'$ band magnitudes and has a peculiar velocity under 100 \kms. However, it has a large offset from the X-ray center, greater than 0.25 $h_{70}^{-1}$ Mpc. There are no other plausible candidates. Identifying either galaxy as the BCG would indicate that RX J0910+5419 has been recently disturbed or not completely formed (\citealt{bird94,GB02}; see also Section \ref{sec:anal} for a discussion of other evidence supporting this conclusion.). In this paper, we choose the brighter, more centrally located galaxy as the BCG. 

\subsubsection{Luminosity-weighted Mean Centers}
\label{sec:LWMC}

We also measure the mean coordinates of all galaxies within the clusters, weighted by the luminosities of the individual galaxies. Galaxies are weighted by their $i'$ luminosities, except in the case of the Cl 1604 supercluster, where we use the F814W luminosities. These measurements are listed in Table \ref{opcentable}. In order to calculate the mean centers, we had to first define which galaxies to include as cluster members. As described in the next section, we chose a circular region with a 1 $h_{70}^{-1}\ $Mpc radius around each cluster, centered on the red galaxy density peaks. Note that 1 $h_{70}^{-1}\ $Mpc is comparable to $r_{500}$ for these clusters \citep[see e.g.,][]{maug11}. The galaxies used for these measurements are the same as those used to calculate velocity dispersions (see Table \ref{specanntab} for the number of galaxies used for each). 

\begin{figure*}
\epsscale{1.2}
\plotone{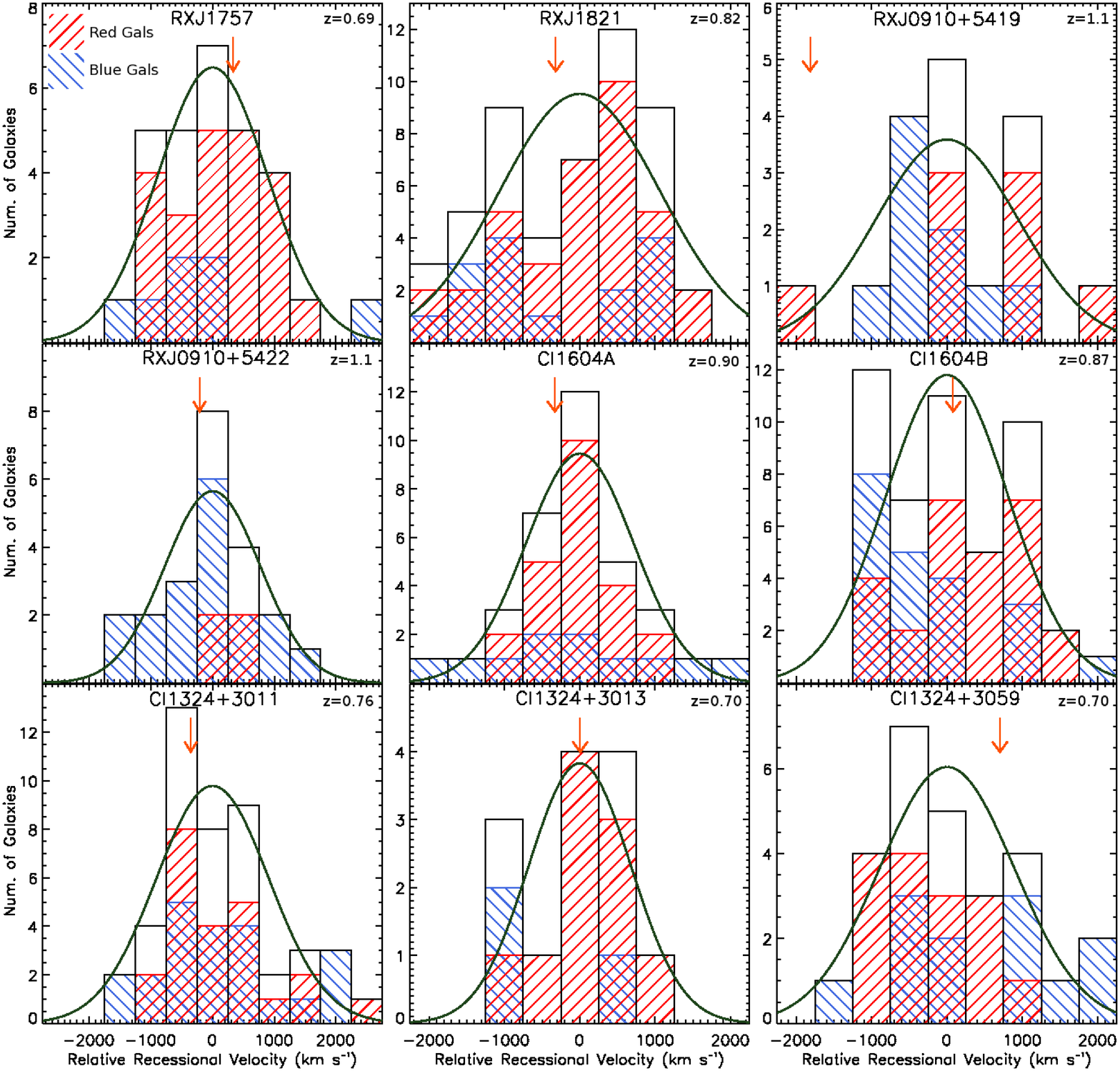}
\caption{\footnotesize{
Velocity histograms are plotted for each cluster, relative to the mean velocity of that cluster. Histograms for the entire cluster population are plotted with unfilled histograms, while velocity histograms of only the red and blue galaxies are plotted with hashed histograms (see legend in upper-left panel). In addition, the solid line is a normal distribution with $\sigma$ equal to the velocity dispersion calculated using the entire galaxy population. The arrows show the velocities of the BCGs. 
}}
\label{velhists}
\end{figure*}

\subsection{Velocity Dispersions}
\label{sec:veldisp}

We measure velocity dispersions following the technique described in \citet{lub02} and \citet{gal05}. For each cluster, we consider galaxies within a $1.0\ h_{70}^{-1}\ $Mpc radius of the red galaxy density peak. While the choice of cluster center could change the galaxies used to compute the dispersions, we choose the red galaxy density peak for consistency with our previous work. Tests using the other possible centroids (described in the previous section) show that the resultant dispersions are consistent within the measurement errors. To illustrate the consistency, we list the velocity dispersions using the BCG centroids in Table \ref{specanntab}.

An initial redshift range is chosen based on visual inspection of each cluster's redshift histogram. Iterative 3$\sigma$-clipping is performed for each cluster, where $\sigma$ is the biweight dispersion as in \citet{beers90}. The final velocity dispersion is taken from the biweight scale estimator after all iterations of the $\sigma$-clipping are complete, and errors are calculated with jackknife confidence intervals. Other measures of the velocity dispersion are consistent within the errors\footnote{The measurements of the velocity dispersions using the biweight scale estimator were always consistent within the errors with gapper measurements, but were sometimes inconsistent with the less robust {\it f}-pseudosigma and median absolute deviation measures.}. Our measurements are presented in Table \ref{specanntab}. Note that some velocity dispersions differ from values published in previous works \citep[see][and references therein]{rum12} due to the addition of new spectroscopic observations and our adoption of a uniform definition of the cluster center. 

Velocity histograms for each cluster are shown in Figure \ref{velhists}. The x-axis represents the velocities relative to the mean recessional velocity of the cluster. A normal distribution, using the velocity dispersion of the cluster, is overplotted with a solid line in each histogram. An arrow is also shown at the velocity of the brightest cluster galaxy (see Section \ref{sec:optgalcen}). 

\subsubsection{Analysis of Red versus Blue Galaxy Populations}
\label{sec:rvb}

While we measured the velocity dispersions of our nine clusters using their entire galaxy populations, studying the blue and red populations separately can yield insights on the dynamics of the cluster \citep[e.g.,][]{ZF93}. In Figure \ref{velhists}, velocity histograms for the red and blue galaxy populations, separated by the lower boundary of the red sequence (see Section \ref{sec:optim}), are shown for those clusters with at least ten galaxies of each color. In a virialized cluster, we would expect the dispersion of the red, relaxed core to differ from that of the bluer infalling populations. This is supported by \citet{ZF93}, who find such differences between late-type and early-type galaxies in low-redshift clusters. When dispersions for the red and blue galaxy populations are similar, it may indicate a younger cluster where the core has not had time to differentiate from the infalling populations. For clusters with sufficient numbers of red and blue galaxies, the dispersions for the two subpopulations are given in Table \ref{DStab}. For most of our clusters, there is a large difference between these values. The two dispersions are most consistent for Cl 1604B, where they differ by only $\sim 10\%$, suggesting this cluster is most likely unrelaxed. 

While differences between the velocity dispersions of red and blue galaxies indicate a relaxed state, differences between the velocity centers of these populations can be a sign of substructure \citep{ZF93}. In Figure \ref{velhists}, we can see that some of the red and blue galaxy populations within a cluster do have differing mean velocities. For example, in Cl 1324+3059, there are six blue galaxies and only one red galaxy with velocities $> 750$ \kms. We can quantify the centers of the red and blue galaxy populations with the biweight location estimator, as defined in \citet{beers90}. The difference between these values for each cluster with enough members is shown in Table \ref{DStab}. We observe both large and small differences, and it is likely that some of these arise by chance, especially in cases with small numbers of galaxies. In a cluster with $X$ confirmed members, $X_B$ of them blue and $X-X_B$ of them red, differences will arise between the velocity centers of the red and blue populations through the process of dividing the galaxies into these two groups. To estimate the significance of the differences, we performed simulations, randomly grouping the entire population of galaxies in each cluster in two groups with $X_B$ and $X-X_B\ $members and measuring the biweight location estimator for both groups. One million trials were performed for each cluster. The percentage of trials where the velocity offsets between the two simulated population exceeded the observed differences between the red and blue populations is given for each cluster in Table \ref{DStab} and is an estimate of the significance of the observed velocity difference\footnote{We also estimated the significance of the velocity center differences through bootstrapping and found very similar results.}. From these results, we can see the 650 \kms\ offset between the red and blue populations in Cl 1604B is significant, with only a 0.8\% chance of arising randomly, while the similar difference in Cl 1324+3059 is moderately significant, with a 9.7\% chance of arising randomly. The difference observed in RX J1821, however, is not significant. These results suggest that substructure or asymmetric infall is present in Cl 1604B and Cl 1324+3059. In the next section, we further explore substructure, for all nine clusters in our sample, with the Dressler-Shectman test.

\begin{deluxetable*}{lcccccccccc}
\tablecaption{Cluster and ICM Properties}
\tablehead{
\colhead{\footnotesize{Cluster}}
& \colhead{\footnotesize{$\left<z\right>$}}
& \colhead{\footnotesize{Num. of}}
& \colhead{\footnotesize{$\sigma_v$\tablenotemark{b}}}
& \colhead{\footnotesize{X-ray}}
& \colhead{\footnotesize{X-ray}}
& \colhead{\footnotesize{Ext.}}
& \colhead{\footnotesize{Net}}
& \colhead{\footnotesize{SNR}}
& \colhead{\footnotesize{Bol.}}
& \colhead{\footnotesize{Gas}} \\
\colhead{\footnotesize{}}
& \colhead{\footnotesize{}}
& \colhead{\footnotesize{Members\tablenotemark{a}}}
& \colhead{\footnotesize{}} 
& \colhead{\footnotesize{Centroid}}
& \colhead{\footnotesize{Centroid}}
& \colhead{\footnotesize{Region}}
& \colhead{\footnotesize{Photon}}
& \colhead{\footnotesize{}}
& \colhead{\footnotesize{X-ray}}
& \colhead{\footnotesize{Temp.\tablenotemark{f}}} \\
\colhead{\footnotesize{}}
& \colhead{\footnotesize{}}
& \colhead{\footnotesize{}}
& \colhead{\footnotesize{}} 
& \colhead{\footnotesize{RA (J2000)}}
& \colhead{\footnotesize{Dec. (J2000)}}
& \colhead{\footnotesize{Radius\tablenotemark{c}}}
& \colhead{\footnotesize{Counts\tablenotemark{d}}}
& \colhead{\footnotesize{}}
& \colhead{\footnotesize{Lum.\tablenotemark{e}}}
& \colhead{\footnotesize{}}
}
\startdata
RX J1821 & 0.818 & 51& 1070$\pm90\phm{0}$ (1090)	& 18\ 21\ 32.3 & +68\ 27\ 57  &  60(450) & 539(645) & \phm{.}19 & 8.8$\pm0.4$ & 5.0$^{+1.0}_{-0.7}$ \\
RX J1757& 0.692 & 29& \phm{1}890$\pm140\phm{1}$ (890) & 17\ 57\ 19.3 & +66\ 31\ 29   &  50(360) & 253(296) & \phm{.}13 & 2.8$\pm0.2$ & 3.8$^{+1.0}_{-0.7}$ \\
RX J0910+5419  & 1.103 & 17& \phm{1}950$\pm190$ (1050)& 09\ 10\ 08.5 & +54\ 18\ 56 &  50(410) & 286(334) & \phm{.}11 & 2.3$\pm0.2$ & 2.5$^{+0.6}_{-0.5}$ \\
RX J0910+5422  & 1.101 & 22& \phm{1}780$\pm140\phm{1}$ (760) & 09\ 10\ 45.0 & +54\ 22\ 07 &  50(410) & 443(462) & \phm{.}16 & 3.6$\pm0.2$ & 4.5$^{+1.1}_{-0.8}$ \\
Cl 1324+3059 & 0.696 & 27 & \phm{1}890$\pm130\phm{1}$ (810)	& 13\ 24\ 49.2 & +30\ 58\ 35 &  80(570) & 151(166) & 6.3 & 1.7$\pm0.2$ & 3.6$^{+3.5}_{-1.6}$ \\
Cl 1324+3011  & 0.755 & 45 & \phm{1}920$\pm120\phm{1}$ (920)	& 13\ 24\ 48.9 & +30\ 11\ 26&  50(370) & 169(210) & 9.3 & 2.6$\pm0.2$ & 3.7$^{+1.4}_{-0.9}$ \\
Cl 1324+3013 & 0.697 & 13  & \phm{1}680$\pm140\phm{1}$ (680)	& 13\ 24\ 20.3 & +30\ 12\ 52 &  80(570) & 157(170) & 6.5 & 1.5$\pm0.2$ &  \tablenotemark{g} \\
Cl 1604A & 0.898 & 34 & \phm{1}720$\pm130\phm{1}$ (720)	& 16\ 04\ 23.5 & +43\ 04\ 39 &  75(580) & 133(122) & 6.5 & 1.9$\pm0.3$ & 3.5$^{+1.8}_{-1.1}$ \\
Cl 1604B & 0.865 & 48  & \phm{1}810$\pm80\phm{0}\phm{1}$ (790)	& 16\ 04\ 26.5 & +43\ 14\ 22 &  50(380)  & \phm{1}\phm{1}69(78)& 4.8  & 1.1$\pm0.3$ & 1.6$^{+0.6}_{-0.5}$\\ 
\enddata
\label{specanntab}
\tablenotetext{a}{\footnotesize{Number of galaxies used to calculate velocity dispersion. See Section \ref{sec:veldisp} and \citet{lub02} or \citet{gal05} for more details of dispersion measurements and criteria for selecting galaxies.}}
\tablenotetext{b}{\footnotesize{Velocity dispersion, in units of \kms, of galaxy cluster members within 1 $h^{-1}_{70}\ $Mpc centered on the red galaxy density peak. The dispersions calculated using the BCG as the center are also shown in parentheses. Errors are given only for the red galaxy density peak centered case.}}
\tablenotetext{c}{\footnotesize{Radius, in arcseconds (kpc), of the region in which the X-ray spectra were extracted.}}
\tablenotetext{d}{\footnotesize{Background-subtracted {\it Chandra} photon counts within the extraction radius (within $r_{500}$). See Section \ref{sec:DE} and the appendix for more details.}}
\tablenotetext{e}{\footnotesize{Rest-frame bolometric X-ray luminosity for the ICM of the given cluster, measured in units of 10$^{44}\ h_{70}^{-2}\ $erg s$^{-1}$, within $r_{500}$.}}
\tablenotetext{f}{\footnotesize{Gas temperature of the ICM measured from the X-ray spectra of the given cluster, in units of keV.}}
\tablenotetext{g}{\footnotesize{Large errors prevented a measurement of the gas temperature for Cl 1324+3013 with any precision.}}
\end{deluxetable*}

\begin{figure*}
\epsscale{1.105}
\plotone{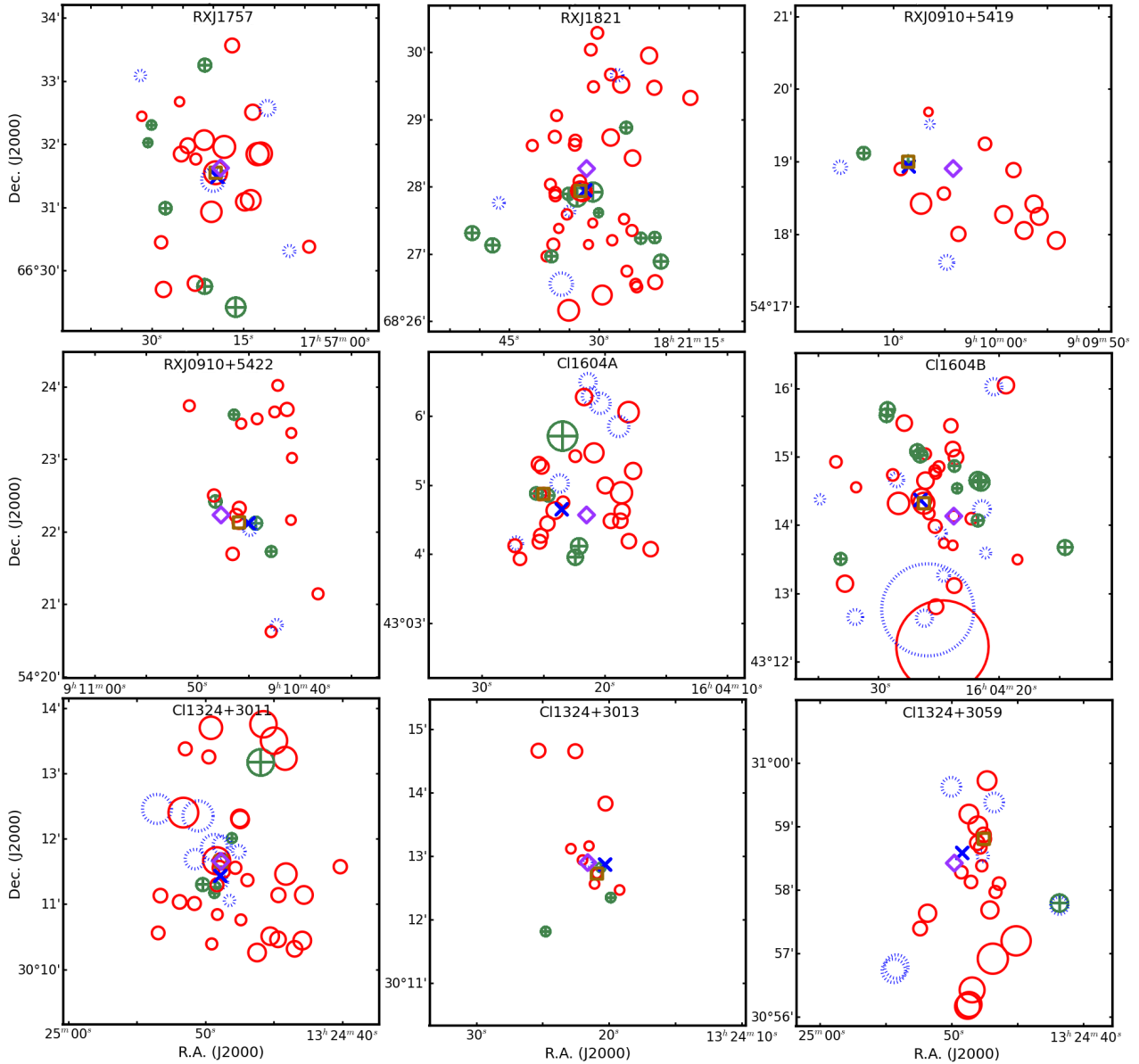}
\caption{\footnotesize{
Shown is a representation of the substructure in each cluster, modeled after Figure 7 of \citet{hall04}. Each galaxy in each cluster within 1 $h_{70}^{-1}\ $Mpc of the red galaxy density peak is plotted, with the size of its representative circle proportional to $e^{\delta}$, where $\delta$ is an output of the D-S tests (see Section \ref{sec:DStests}). Galaxies represented by solid, unfilled circles have velocities within $\sigma_v$ of the mean velocity of the cluster, while those represented by dashed circles and crossed circles have velocities above and below this range, respectively. The centers of X-ray emission, the BCGs, and the red galaxy density peaks are represented by X's, squares, and diamonds, respectively. Each panel is 2.2 $h_{70}^{-1}$\ Mpc on a side.  
}}
\label{DSfig}
\end{figure*}

\begin{deluxetable*}{lcccccccc}
\tablecaption{Results of Substructure and Dynamical State Tests}
\tablehead{
\colhead{\footnotesize{Structure}}
 & \colhead{\footnotesize{$\Delta$\tablenotemark{a}}}
 & \colhead{\footnotesize{$P$\tablenotemark{b}}}
 & \colhead{\footnotesize{Num. of Blue}}
 & \colhead{\footnotesize{Blue Vel.}}
 & \colhead{\footnotesize{Num. of Red}}
 & \colhead{\footnotesize{Red Vel.}}
 & \colhead{\footnotesize{Red/Blue}}
 & \colhead{\footnotesize{Prob.\tablenotemark{e}}}\\
  \colhead{\footnotesize{}}
 & \colhead{\footnotesize{}}
 & \colhead{\footnotesize{}}
 & \colhead{\footnotesize{Members}}
 & \colhead{\footnotesize{Dispersion\tablenotemark{c}}}
 & \colhead{\footnotesize{Members}}
 & \colhead{\footnotesize{Dispersion\tablenotemark{c}}}
 & \colhead{\footnotesize{Vel. Offset\tablenotemark{d}}}
 & \colhead{\footnotesize{(\%)}}\\
  \colhead{\footnotesize{}}
 & \colhead{\footnotesize{}}
 & \colhead{\footnotesize{}}
 & \colhead{\footnotesize{}}
 & \colhead{\footnotesize{(\kms)}}
 & \colhead{\footnotesize{}}
 & \colhead{\footnotesize{(\kms)}}
 & \colhead{\footnotesize{(\kms)}}
 & \colhead{\footnotesize{}}
}
\startdata
RX J1821	& 41.1 & 0.94 & 15 & 1160 & 36 & 950 & 640 & 16\\
RX J1757	& 34.2 & 0.16 & \phm{1}7 & \nodata & 22 & \nodata & \nodata & \nodata \\
RX J0910+5419 & 17.6 & 0.17 & \phm{1}9 & \nodata & \phm{1}8 & \nodata & \nodata & \nodata \\
RX J0910+5422 & 14.5 & 0.72 & 18 & \nodata & \phm{1}4 & \nodata & \nodata & \nodata \\
Cl 1324+3059	& 40.6 & 0.04 & 12 & 1040 & 15 & 620 & 640 & 9.7\\
Cl 1324+3011	& 60.6 & 0.15 & 22 & 1010 & 23 & 820 & \phm{6}40 & 89\\
Cl 1324+3013	& \phm{1}6.2& 0.47 & \phm{1}3 & \nodata & 10 & \nodata & \nodata & \nodata \\
Cl 1604A	& 43.8 & 0.17 & 11 & 1210 & 23 & 500 & \phm{6}40 & 88\\
Cl 1604B	& 54.0 & 0.28 & 21 & \phm{1}770 & 27 & 710 & 650 & 0.8
\enddata
\label{DStab}
\tablenotetext{a}{\footnotesize{Diagnostic output of Dressler-Shectman tests. See Section \ref{sec:DStests} for details.
}}
\tablenotetext{b}{\footnotesize{Estimate of the probability, given as a fraction from zero to one, that the cluster does not contain substructure, derived from DS tests using Monte Carlo simulations.}}
\tablenotetext{c}{\footnotesize{Velocity dispersion calculated using only blue/red cluster members. See Section \ref{sec:rvb} for details. These values were not calculated for clusters with fewer than ten members in either the red or blue populations. }}
\tablenotetext{d}{\footnotesize{Difference between velocity centers, measured using the biweight location estimator, of the red and blue populations of each cluster (see Section \ref{sec:rvb} for details). These values were not calculated for clusters with fewer than ten members in either the red or blue populations.}}
\tablenotetext{e}{\footnotesize{Probability that a velocity difference as large as that observed in the previous column would arise by chance. See Section \ref{sec:rvb} for derivation of these values.
}}
\end{deluxetable*}

\subsection{Dressler-Shectman Tests of Substructure}
\label{sec:DStests}

As an alternate means to detect substructure, we apply the Dressler-Shectman (D-S) test, which employs both the spatial positions and velocities of galaxies \citep{DS88,hall04}. The D-S test uses the statistic \begin{equation}
\delta^2 = \frac{11}{\sigma_v^2}\left[\left(v_{loc} - \bar{v}\right)^2+\left(\sigma_{loc}-\sigma_v\right)^2\right]
\end{equation}
\label{DSeq}
where $\bar{v}$ and $\sigma_v$ are the mean velocity and velocity dispersion of the cluster, respectively. The variables $\delta$, $v_{loc}$, and $\sigma_{loc}$ are calculated for an individual galaxy in the cluster. The mean velocity and velocity dispersion of that galaxy and its ten nearest neighbours within the cluster are represented by $v_{loc}$ and $\sigma_{loc}$, respectively. As a measure of the total substructure present in a cluster, \citet{DS88} use $\Delta$, the sum of the $\delta$-values of each galaxy. $\Delta$ has a distribution like $\chi^2$, and its expected value is on the order of the number of galaxies in the cluster. 

The results of the D-S tests are displayed in Table \ref{DStab}. To estimate the significance of $\Delta$ for each cluster, we employed a series of Monte Carlo (MC) tests, using the method of \citet{hall04}. For each cluster, we randomly shuffled the velocities among all galaxies and recalculated $\Delta$. We carried out 1000 trials and measured the fraction $P$ of trial values of $\Delta$ that were greater than the actual measured value of $\Delta$. 

While $\Delta$ is a measure of the overall amount of substructure in a cluster, it does not provide information on where the substructure is located. For that purpose, we created Figure \ref{DSfig}, modeled after Figure 7 of \citet{hall04}. Each circle represents one galaxy in the cluster, and the size of the circle is proportional to $e^{\delta}$. Substructure will be apparent by many large circles in close proximity. 

From our D-S tests, we can see that many of the clusters in our sample likely contain substructure. However, the results are not particularly significant, with only Cl 1324+3059 showing substructure at a $>90\%$ confidence level. These findings reinforce our results from the previous section that also indicated that this cluster contains substructure. The findings in the prior section also suggested that Cl 1604B had substructure or asymmetric infall, but the D-S test does not indicate substructure at high significance. This discrepancy is not unprecedented \citep{DS88}, and it may mean that the D-S test has failed in this case. From Figure \ref{velhists}, we can see that differences between the blue and red galaxy velocity dispersions in Cl 1604 arises because of a larger number of blue members with lower velocities. In Figure \ref{DSfig}, these galaxies are shown with crossed circles. Examining this plot, we can see the crossed circles appear to have some clustering, with three pairs of crossed circles with very small separations. While the D-S statistic did not imply significant substructure, the figure does indicate a complex dynamical structure. Due to incomplete spectroscopy, it is possible that substructure could have been missed in some clusters in our sample. In Section \ref{sec:anal}, we analyze how these substructure measures relate to other properties of the clusters. 

\begin{figure*}[t]
\epsscale{1.1}
\plotone{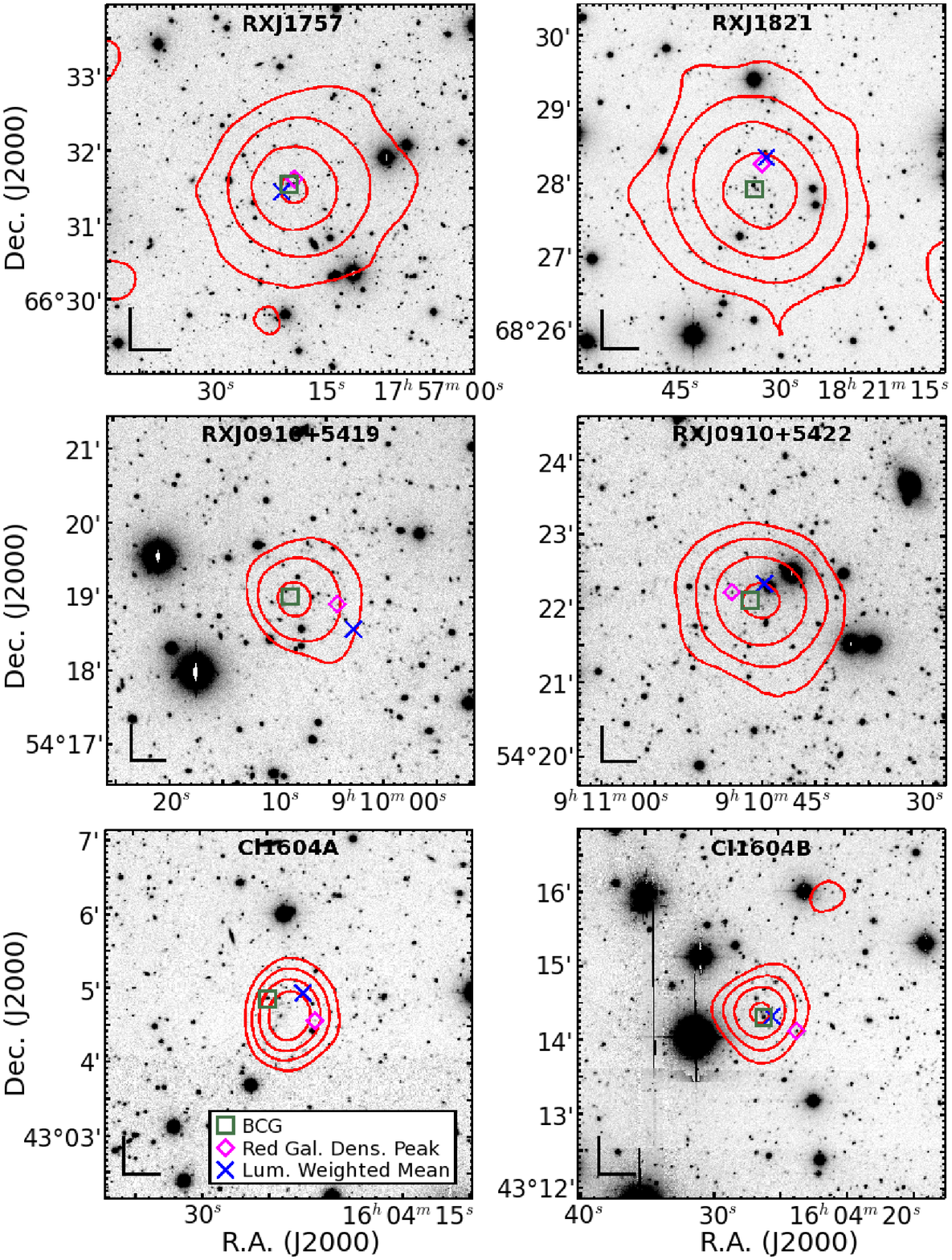}
\caption{\footnotesize{
Smoothed X-ray contours for six clusters from our sample overlaid on optical $i'$-band images. Also shown are positions of the BCGs (squares), red galaxy density peaks (diamonds), and luminosity-weighted mean positions (X's). Each image is centered on the peak of the diffuse emission and is 5$' \times $5$'$. Note that RX J0910+5419 lies near the corner of a {\it Chandra} chip, which has slightly perturbed the X-ray contours. The contour levels correspond to the following levels of significance above the background for the following clusters: RX J1757 - 3$\sigma$, 6$\sigma$, 10$\sigma$, 13$\sigma$; RX J1821 - 3$\sigma$, 6$\sigma$, 10$\sigma$, 15$\sigma$; RX J0910+5419 - 3$\sigma$, 6$\sigma$, 10$\sigma$; RX J0910+5422 - 3$\sigma$, 6$\sigma$, 10$\sigma$, 15$\sigma$; Cl 1604A - 3$\sigma$, 4$\sigma$, 5$\sigma$, 6$\sigma$; Cl 1604B - 3$\sigma$, 4$\sigma$, 5$\sigma$, 6$\sigma$. Here, $\sigma$ refers to a Poissonian error. X-ray contours for the remaining three clusters in our sample are plotted in Figure \ref{consfig2}. The angle in the lower left of each image has sides of length 0.25 Mpc at the redshift of the respective cluster. 
}}
\label{consfig1}
\end{figure*}

\begin{figure*}[t]
\epsscale{1.1}
\plotone{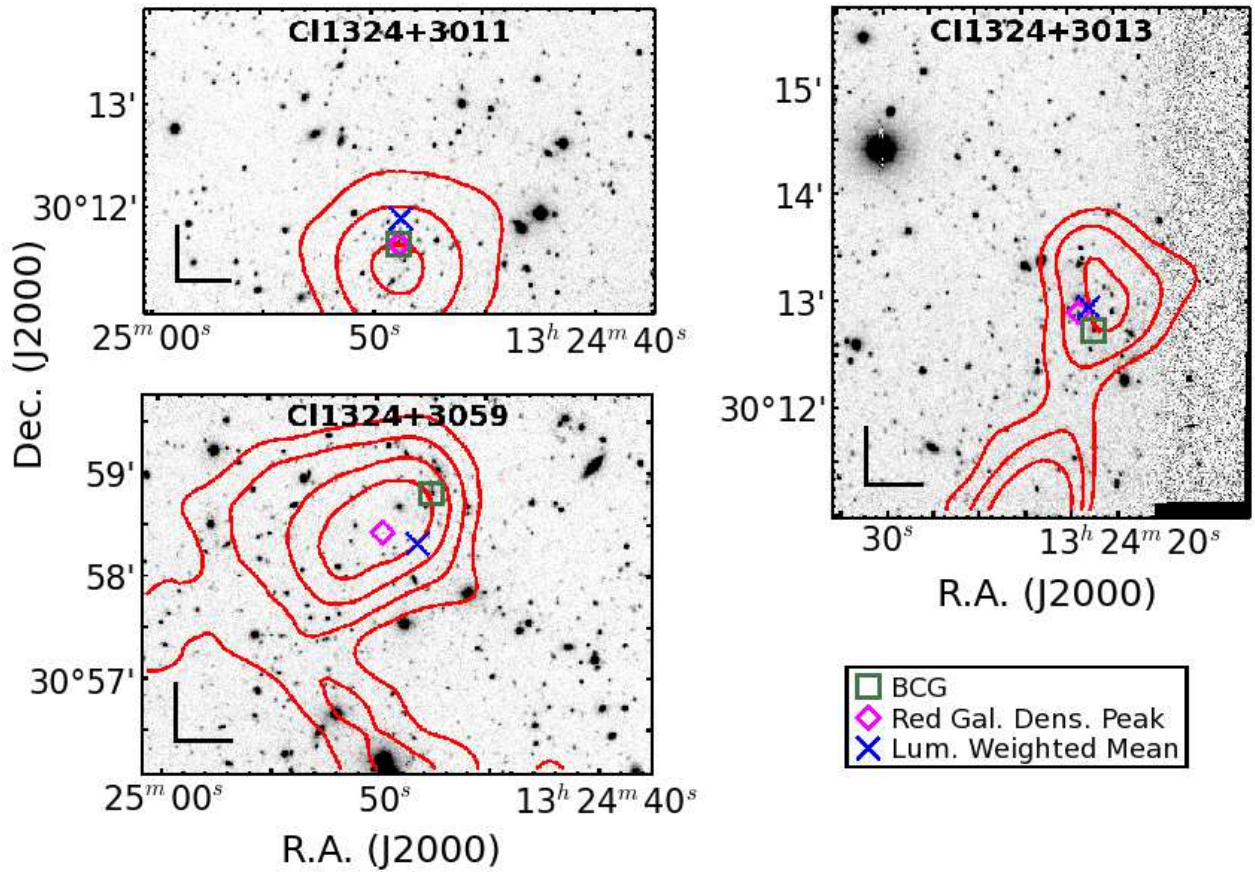}
\caption{\footnotesize{
Smoothed X-ray contours for the three clusters in the Cl 1324 supercluster overlaid on optical $i'$-band images. Also shown are positions of the BCGs (squares), red galaxy density peaks (diamonds), and luminosity-weighted mean positions (X's). Each image is centered on the peak of the diffuse emission. While images in Figure \ref{consfig1} were each 5$' \times $5$'$, the three clusters from Cl 1324 were all near the edge of LFC imaging, so these plots are subsequently smaller in size. Note that there is an extended X-ray source very close to Cl 1324+3013 to the south, which is a foreground cluster \citep{GY05}. The contour levels correspond to the following levels of significance above the background for the three clusters: Cl 1324+3011 - 3$\sigma$, 6$\sigma$, 9$\sigma$; Cl 1324+3013 - 3$\sigma$, 4$\sigma$, 5$\sigma$; Cl 1324+3059 - 3$\sigma$, 4$\sigma$, 5$\sigma$, 6$\sigma$. Here, $\sigma$ refers to a Poissonian error. The angle in the lower left of each image has sides of length 0.25 Mpc at the redshift of the respective cluster. 
}}
\label{consfig2}
\end{figure*}

\subsection{Diffuse X-ray Emission}
\label{sec:DE}

To search for diffuse X-ray emission, {\it Chandra} images, with point sources removed, were smoothed using two dimensional, azimuthally symmetric beta models:
\begin{equation}
f\left(r\right) = A\left(1+\frac{r^2}{r_c^2}\right)^{-3\beta+1/2}
\end{equation}
We used $\beta = 2/3$ and a core radius of $r_c = 180$ kpc, which are typical values for clusters\footnote{Smoothing was also performed adaptively with a gaussian kernel for comparison. We found the different values for the X-ray centroid produced by the two different smoothing techniques to be small ($< 2.5$\ arcseconds, except for the two most irregular clusters.) compared to the distances between optical and X-ray centroids.} \citep[see e.g.,][]{AE99,ett04,maug06,hicks08}. Centroids for the diffuse emission peaks are listed in Table \ref{specanntab}. The member groups and clusters in Cl 1604 and Cl 1324 that are not listed in Table \ref{specanntab} have no observed diffuse emission.

Smoothed X-ray contours for each cluster, overlaid on optical $i'$ images, are displayed in Figures \ref{consfig1} and \ref{consfig2}. The images are $5^\prime$ (or $\sim 2$ Mpc) on each side, except for the clusters in the Cl 1324 supercluster, which were too close to the edge of our optical imaging for this sizing. Contour levels are listed in the appropriate figure captions. While we can visually look for asymmetries, it is important to note that our data are insufficiently deep for precise analysis of the X-ray contours. Keeping this in mind, asymmetries can be observed in the X-ray contours of RX J0910+5419, Cl 1604A, Cl 1604B, Cl 1324+3013 and Cl 1324+3059, although the asymmetry in RX J0910+5419 could be the result of its proximity to the {\it Chandra} chipgap. For any of these systems, the asymmetry could be caused by a recent merger, an infalling population, or other significant substructure. Also of note is the peak visible to the south of Cl 1324+3013, which is a foreground cluster \citep{GY05}. Although the clusters lie in close proximity, they probably do not significantly affect each others' X-ray contours. 

To compute gas temperatures for the clusters, spectra were obtained for each emission peak using the CIAO tool {\it specextract}. One-dimensional surface brightness profiles were measured around each peak to determine the region to use for extraction of the spectrum. For each profile, the radius where the surface brightness reached the background level was determined, and the spectrum was measured in a circular region within this radius. The radii of these regions are listed in Table \ref{specanntab}.  A background spectrum was also extracted in a surrounding annulus and then subtracted. The outer radii of the background regions were twice as large as the spectrum extraction regions, except for clusters too close to the edge of the chip. For more information on how these regions were determined, see the appendix. 

The spectra were fit to a Raymond-Smith thermal plasma model \citep{RS77}, with the absorption model of \citet{MM83}, which we chose for consistency with previous work. Fitting, as well as error determination, was accomplished using the Sherpa tool. In our fits, we assumed $Z = 0.3Z_{\odot}$ \citep{ES91}\footnote{While $Z = 0.3Z_{\odot}$\ is commonly used in the literature, we found that varying the metallicity did not have a large effect on fitting, with temperature variations much lower than the overall errors.}. Galactic neutral hydrogen column densities were calculated at the aim point of each observation using the COLDEN tool from the {\it Chandra} proposal toolkit, using the dataset of \citet{DicLock90}. Fits were performed on the 0.5-8.0 keV energy range, using $\chi^2$ statistics. Because of the low number of counts, spectra were grouped to include a minimum of 20 counts per bin. Bin sizes with a variable number of minimum counts were tested to ensure our choice did not significantly affect the measured temperature. The results of the fits are shown in Table \ref{specanntab}. 

Note that no temperature could be measured for Cl 1324+3013, although diffuse emission was detected from the cluster. Errors on the fit to the ICM for this cluster, calculated using the same Sherpa fitting procedure, were too high to determine a temperature with any precision. With longer exposures on the clusters, it may be possible to measure a temperature for Cl 1324+3013, as well as investigate the contour asymmetries in more depth. 

\begin{figure}
\epsscale{1.2}
\plotone{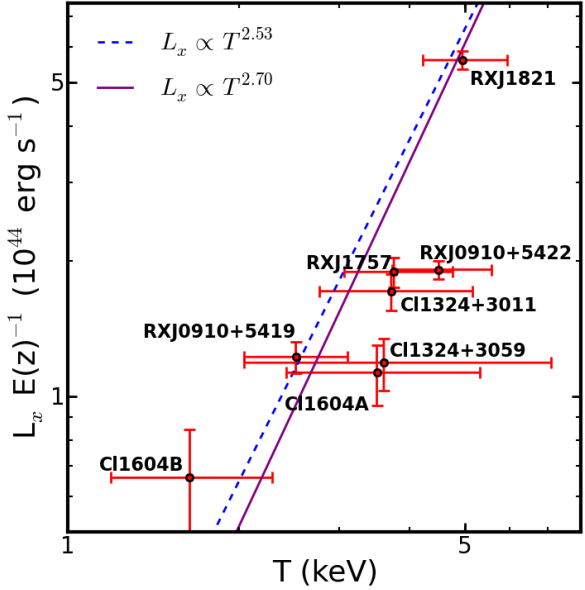}
\caption{\footnotesize{
Bolometric rest-frame X-ray luminosities of cluster ICM gas plotted versus gas temperatures. The local $L_x$-$T$ relationships of \citet{pratt09} and \citet{rei11} are shown with solid and dashed lines, respectively.
}}
\label{LxTfig}
\end{figure}

Since net photon counts were measured in regions whose sizes varied from cluster to cluster, we made an extrapolation out to $r_{500} = 2\sigma_v/\left[\sqrt{500}\ H(z)\right]$ to compute a luminosity that is easier to compare between our clusters and those in other studies. Note that $r_{500}$ is only one to two times the extraction radius for our sample. This extrapolation was accomplished using beta models that were fit to our data (see the appendix for more details). In these fits, the core radii of the models were varied and the errors from the fit were propagated through our subsequent calculations. Since the fitted Raymond-Smith profiles for the diffuse cluster emission provide a relation between photon count rates and flux in a given energy range, we used them to convert our measured total photon count rates to fluxes in the different energy bands, correcting for galactic absorption by HI at the same time. For a source at redshift $z$, observed fluxes in the 0.5 to 2.0 keV range were translated to observer-frame fluxes in the energy range 0.5/$\left(1+z\right)$ to 2.0/$\left(1+z\right)$ using the Raymond-Smith models. These fluxes were then multiplied by $4\pi D_L^2\ $to recover the luminosity emitted in the rest frame 0.5 to 2.0 keV energy range. Using the Raymond-Smith spectral models, as in \citet{koc09a}, these rest-frame luminosities were extrapolated to the bolometric rest-frame luminosities listed in Table \ref{specanntab}.

\section{Cluster Scaling Relations}
\label{sec:scalrel}

In this section, we examine the relationships between the diffuse gas temperatures, the bolometric ICM X-ray luminosities, and the velocity dispersions of the clusters and compare them to local scaling relations.

\begin{deluxetable}{lccc}
\tablecaption{Offsets From Scaling Relations}
\tablehead{
\colhead{\footnotesize{Structure}}
 & \colhead{\footnotesize{Offset From}}
 & \colhead{\footnotesize{Offset From}}
 & \colhead{\footnotesize{Offset From}}\\
  \colhead{\footnotesize{}}
 & \colhead{\footnotesize{$L_x$-$T$ Rel.\tablenotemark{a}}}
 & \colhead{\footnotesize{$\sigma_v$-$T$ Rel.\tablenotemark{a}}}
 & \colhead{\footnotesize{$L_x$-$\sigma_v$ Rel.\tablenotemark{a}}}
}
\startdata
RX J1821 & 0.1 & 1.4 & 1.7 \\ 
RX J1757 & 0.7 & 0.9 & 1.0 \\ 
RX J0910+5422 & 1.6 & 0.3 & 0.3 \\ 
RX J0910+5419 & 0.4 & 1.8 & 1.4 \\
Cl 1324+3059 & 0.6 & 0.4 & 1.6 \\ 
Cl 1324+3011 & 0.6 & 0.9 & 1.7 \\ 
Cl 1324+3013 & \nodata & \nodata & 0.1 \\ 
Cl 1604A & 0.8 & 0.1 & 0.0 \\ 
Cl 1604B & 0.8 & 3.1 & 2.1 
\enddata
\label{offsetstab}
\tablenotetext{a}{\footnotesize{Minimum-distance offsets from respective scaling relations. Offsets are given in units of $\sigma$, assuming Gaussian errors. Note that no temperature was measured for Cl 1324+3013. See Section \ref{sec:scalrel}.
}}
\end{deluxetable}

\subsection{$L_x$-$T$}

If the ICM gas is subjected only to gravity, and thus, gravitational collapse is the only source of heating, we would expect the X-ray luminosity to scale with the temperature as $L_x \propto T^2$, with the photon source being bremsstrahlung emission \citep{kaiser86}. Additionally, the proportionality coefficient for this relationship is expected to evolve with redshift as $E\left(z\right)$ \citep{krav06}, with  \begin{equation}
E\left(z\right) = \left[\Omega_m\left(1+z\right)^3 + \left(1-\Omega_m-\Omega_{\Lambda}\right)\left(1+z\right)^2 + \Omega_{\Lambda}\right]^{1/2}
\end{equation}

\begin{figure}
\epsscale{1.2}
\plotone{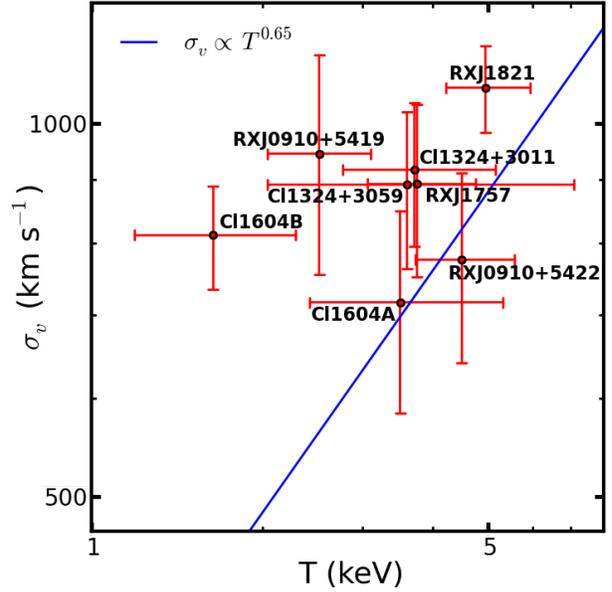}
\caption{\footnotesize{
Velocity dispersions of galaxy cluster members are plotted versus ICM gas temperatures. Also plotted is the empirical relation between the two properties of \citet{XW00} using a sample of low redshift clusters.
}}
\label{sigTfig}
\end{figure}

A number of studies have found that clusters do not follow the $L_x \propto T^2$ relation, with low-redshift studies finding a steeper relation, closer to $L_x \propto T^3$ \citep{mark98,AE99,XW00,vik02}. This result implies an injection of energy from another source besides gravitational heating, such as AGN. Studies have also found that it does not evolve with redshift in a self-similar fashion \citep[e.g.,][]{rei11}. We examine this evolution in more detail in Section \ref{sec:redev}.

In Figure \ref{LxTfig}, bolometric X-ray luminosities (measured within $r_{500}$) for the clusters studied in this paper are plotted against their ICM gas temperatures, corrected for self-similar evolution. The local $L_x$-$T$ relationships of \citet{pratt09} and \citet{rei11} are plotted, which follow $L_x \propto T^{2.79}$ and $L_x \propto T^{2.53}$, respectively. We can see that, except for RX J0910+5422, all the clusters appear to be consistent with the local scaling relations. Minimum-distance offsets from the \citet{pratt09} relation are shown in Table \ref{offsetstab} for each cluster, in units of $\sigma$, assuming Gaussian errors. From this table, we can see that RX J0910+5422 has the largest offset, although at a marginal significance of $1.6\sigma$.  An offset from the relation could indicate a recent merger or that the cluster is still in the process of forming. 

\subsection{$\sigma_v$-$T$}

The $\sigma_v$-$T$ relation relates the optical and X-ray properties of the clusters. If the ICM follows the same dynamics as the cluster galaxies, assuming virialization, we would expect the X-ray luminosities of the gas and the velocity dispersion of the galaxies to be related by $\sigma_v \propto T^{1/2}$. As with the $L_x$-$T$ relationship, local studies have found a deviation from this prediction, with a slightly higher power of $T$ \citep[e.g.,][]{XW00,horner01}, which would imply non-gravitational sources of heating.

In Figure \ref{sigTfig}, we plot the velocity dispersions against the ICM temperatures for our clusters. An empirical relationship between the two quantities is also plotted, from the local study of \citet{XW00}, with $\sigma_v \propto T^{0.65}$. Almost all of the clusters in our sample are consistent with the local scaling relation. We can see from Table \ref{offsetstab} that only three clusters are offset from the local relation by more than 1$\sigma$. Cl 1604B has the most substantial offset, over 3$\sigma$, followed by RX J0910+5419 and RX J1821. Although these results are only significant for Cl 1604B, they may indicate that these clusters have temperatures that are lower than those for virialized clusters with the same velocity dispersions, which may be because they are still gravitationally heating (e.g., the gas can be distributed among many smaller, cooler subclumps or incomplete relaxation can result in substantial substructure; \citealt{frenk96,valt04,cast11}). Alternatively, the dispersions could be inflated due to infalling galaxies, a filament oriented along the line of sight, or a recent merger \citep[see e.g.,][]{bow97,gioia99}. While our substructure tests from the previous section showed that Cl 1604B likely has infall or substructure, no significant substructure was detected in the other two clusters.

\begin{figure}
\epsscale{1.2}
\plotone{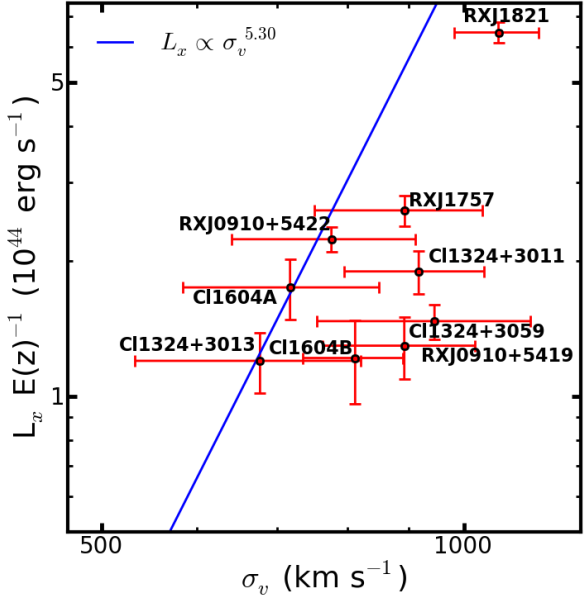}
\caption{\footnotesize{
Velocity dispersions of galaxy cluster members are plotted versus the bolometric rest-frame X-ray luminosities of the ICM gas, extrapolated to infinity for comparison to the local empiral relation of \citet{XW00}, which is shown with a line.
}}
\label{sigLxfig}
\end{figure}

\subsection{$L_x$-$\sigma_v$}

Similarly to the $\sigma_v$-$T$ relation, the $L_x$-$\sigma_v$ relation relates the X-ray and optical properties of a cluster. If one assumes the purely gravitational relations $L_x \propto T^2$ and $\sigma_v \propto T^{1/2}$, we would expect $L_x \propto \sigma_v^4$. This is equivalent to assuming that the galaxies and the ICM are in virial equilibrium and that the total gas mass is proportional to the virial mass of the cluster \citep{QM82}. As has been found for the other two relations, the $L_x$-$\sigma_v$ relation deviates from $L_x \propto \sigma_v^4$ in local studies, with a higher power of $\sigma_v$ \citep[e.g.,][]{XW00,horner01}. Once again, this could be caused by non-gravitational heating.

Compared to relations between different X-ray properties of clusters, there is a lack of studies on relations between their optical and X-ray properties. Lacking a recent relation between $L_{500}$\ and $\sigma_v$, we have chosen to compare to the local $L_x$-$\sigma_v$ relation of \citet{XW00}, which uses X-ray luminosities extrapolated to an infinite radius from their fitted beta models. As described in Section \ref{sec:DE}, we have, therefore, done the same for our bolometric X-ray luminosities for this comparison. In Figure \ref{sigLxfig}, we plot the results\footnote{Note that Cl 1324+3013 is plotted in Figure \ref{sigLxfig}, but not Figures \ref{LxTfig} or \ref{sigTfig}, since we were able to measure an X-ray luminosity and a velocity dispersion for the cluster, but no reliable temperature.}, along with the best-fit line from the local relation of \citet{XW00}.


We can see that some of the clusters are consistent with the local scaling relation, while others are offset below the relation. Cl 1604B, RX J1821, Cl 1324+3011 and Cl 1324+3059 have the most significant offsets, although only Cl 1604B is offset by more than 2$\sigma$. These clusters could be underluminous compared to virialized clusters with the same velocity dispersion. These results could mean that these clusters are young and unrelaxed and have not built up a significant ICM. Once again, the cluster velocity dispersions could be inflated for other reasons, such as from the presence of infalling galaxies, a filament along the line of sight, or a recent merger. This is more likely to be the case for Cl 1324+3011, Cl 1324+3059, and RX J1821, which have higher velocity dispersions than expected from the local scaling relations, but were consistent with the $L_x$-$T$ and $\sigma_v$-$T$ relations. In Section \ref{sec:anal}, we use additional diagnostics to better evaluate the dynamical states of the clusters. 

\subsection{Redshift Evolution of Scaling Relations}
\label{sec:redev}

\begin{figure}
\epsscale{1.2}
\plotone{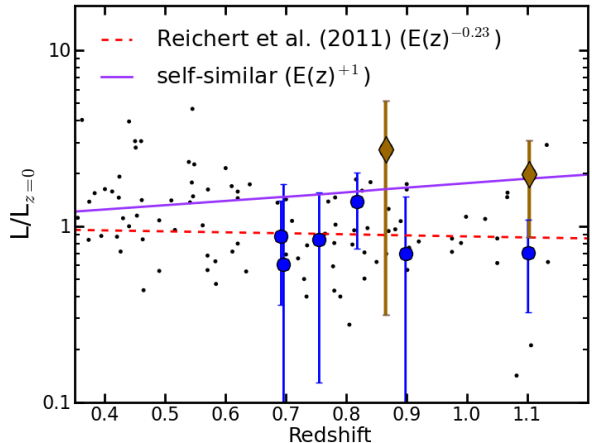}
\caption{\footnotesize{Measured bolometric luminosities for each cluster are plotted divided by $L_{z=0}\left(T\right)$, the luminosity determined from the local $L_x$-$T$ relation for virialized clusters at the temperature for that cluster. Also plotted are expected relations between $L_x/L_{z=0}$ and $z$ assuming self-similarity and from \citet{rei11}. Clusters thought to be unrelaxed are shown with diamonds while the others are shown with circles. Small points represent clusters from the samples gathered in \citet{rei11}.
}}
\label{LxTevolfig}
\end{figure}

A number of studies have examined the evolution of the $L_x$-$T$ relation for viriliazed clusters with redshift. Recently, \citet{rei11} have compiled work from the literature into a large sample of clusters across a redshift range from about $z=0.3$ to 1.5. While self-similarity predicts that $L_x \propto E\left(z\right)^{\alpha}$, with $\alpha=+1$, \citet{rei11} found $\alpha=-0.23^{+0.12}_{-0.62}$. Despite our small sample size, which is unsuitable for fitting the redshift evolution of the $L_x$-$T$ relation, we can compare our data with the relation of \citet{rei11}. We can also determine if our sample is consistent with self-similarity, which other studies have found in this redshift range \citep[e.g., ][]{bran07}.

Plotted in Figure \ref{LxTevolfig} are the bolometric X-ray luminosities (within $R_{500}$; see Section \ref{sec:DE}) for each cluster divided by $L_{z=0}\left(T\right)$, the luminosity predicted from the local $L_x$-$T$ scaling relation used by \citet{rei11}. In this way, we can single out the part of the relation that evolves with redshift. We have plotted all eight clusters for which we have temperature measurements. The two clusters that are likely unrelaxed (Cl 1604B and RX J0910+5419; see Section \ref{sec:anal}) are shown with diamonds, while those considered virialized are shown with circles. In addition, clusters from the sample used in \citet{rei11} are shown with small points. The solid line shows the expected relation assuming self-similarity, while the dashed line is the relation from \citet{rei11}. We can see that, except for RX J0910+5422, at $z\approx1.1$, which is marginally consistent with only the \citet{rei11} relation, the clusters in our sample are within 1.25$\sigma$\ of either line. However, all of the clusters that are more consistent with relaxation are below the line representing self-similar evolution. This suggests a redshift evolution of the $L_x$-$T$ relation that is characterized by $\alpha < 1$. 

\section{Analysis}
\label{sec:anal}

We found in the previous section that most of the clusters in our sample are consistent with local scaling relations between $L_x$, $T$, and $\sigma_v$. The offsets of each cluster from the three scaling relations are shown in Table \ref{offsetstab}, in units of $\sigma$, assuming Gaussian errors. Six of the nine clusters are consistent with at least two of the three local relations or, in the case of Cl 1324+3013, are consistent with the one available relation.  We can see that RX J0910+5422, Cl 1324+3011, and Cl 1324+3059 are 
inconsistent with only one relation, although by less than 2$\sigma$. RX J1821 and RX J0910+5419 were modestly inconsistent with two of the relations. While consistent with the $L_x$-$T$ relation, Cl 1604B was the most offset from the other two relations, with offsets of 2.1$\sigma$ and 3.1$\sigma$. 

These results imply that Cl 1604B is the most likely cluster to be unrelaxed. The modest deviations 
from the scaling relations observed for the two Cl 1324 clusters and RX J1821, on the other hand, could be explained by inflated velocity dispersions. Other diagnostics support this claim. As discussed in Section \ref{sec:rvb}, large differences between the velocity dispersions of red and blue galaxy populations can indicate a relaxed state for a cluster. For the Cl 1324 clusters and RX J1821, we do observe significant differences between the dispersions for red and blue galaxies. Also, our D-S test along with the difference between the velocity centers of the red and blue galaxy populations show that Cl 1324+3059 is the most likely cluster to contain substructure. This result could indicate a recent merger or an infalling population, associated with e.g., a filamentary structure, which could be the cause of the inflated dispersion. While the D-S test for RX J1821 does not yield a high probability that substructure exists, a small group has been observed to the south \citep{lub09}, which is located outside the 1 $h_{70}^{-1}$ Mpc radius used in our D-S tests. These results provide evidence of infalling populations inflating the velocity dispersions for both Cl 1324+3059 and RX J1821. Also, Cl 1324+3059 has one of the most asymmetric X-ray contours in our sample. This evidence suggests a recent merger or more evidence of a filament or other infalling population, which is supported by the large offset between the cluster's BCG and X-ray center. In either case, the evidence suggests a disturbed, unrelaxed cluster, in contradiction to the relaxed state suggested by the red versus blue velocity dispersions. It is possible that the cluster is mostly relaxed, with the bulk of the baryons in equilibrium, but we would require many more spectroscopic redshifts and a detailed X-ray temperature map to confirm this. Altogether, the evidence suggests that the other two clusters, Cl 1324+3011 and RX J1821, are likely to be relaxed but with inflated velocity dispersions that create modest inconsistencies with some scaling relations. 

While this evidence suggests inflated velocity dispersions for some clusters, there is little to suggest the same for Cl 1604B and RX J0910+5419, even though the latter had similar offsets from the scaling relations compared to RX J1821. The dispersions of red and blue galaxies in Cl 1604B are within 10\% of each other, 770 \kms\ versus 710 \kms, the smallest difference measured by a factor of two, suggestive of an unrelaxed cluster. In addition, the velocity centers of the red and blue galaxies differ significantly, which probably indicates some form of substructure in the cluster, although this is not supported by the D-S test. Unfortunately, the small number of confirmed members in RX J0910+5419, as well as three of our other clusters, prevented such measurements for comparison. However, the BCG in RX J0910+5419 has a large peculiar velocity which could indicate a recent merger\footnote{It is somewhat ambiguous which galaxy in RX J0910+5419 is the BCG. However, the other BCG candidate in the cluster has a large offset on the sky from the X-ray center, which has similar implications regarding the dynamical state of the system. See Section \ref{sec:BCG}.} \citep{bird94,GB02}. While the BCG in Cl 1604B does not have a large physical or velocity offset, which would have provided further evidence against virialization, Cl 1604B and RX J0910+5419 have the largest offsets between the red galaxy density peak and X-ray emission centroid of all the clusters in our sample. This result could imply that these clusters are still in the process of forming. The D-S tests for these two clusters do not provide good evidence for substructure, and there is no clear evidence that the velocity dispersions for Cl 1604B and RX J0910+5419 are inflated. The evidence suggests that Cl 1604B and RX J0910+5419 are likely to be unrelaxed clusters.

Three clusters in our sample (Cl 1324+3013, RX J1757, and Cl 1604A) are consistent with all scaling relations while RX J0910+5422 is only marginally inconsistent with the $L_x$-$T$ relation. While relaxed clusters are expected to fall along these relations, it is possible for unrelaxed clusters to do so as well \citep[see e.g.,][]{maug11}. As for the other clusters, we can use additional diagnostics to further probe their dynamical states. Unfortunately, of these four clusters, only Cl 1604A has enough spectroscopically confirmed members to accurately determine red and blue galaxy velocity dispersions separately. As expected for a relaxed cluster, we find significantly different velocity dispersions and consistent mean velocities for the red and blue galaxy populations in Cl 1604A. Three of the four clusters have undisturbed X-ray contours, with Cl 1324+3013 being the sole exception. In the case of Cl 1324+3013, the highly asymmetric contour could indicate a recent merger or that the cluster is still in the process of forming. Since we were unable to measure a temperature for this cluster, we do not know if it is consistent with two of the three scaling relations studied. The data that we have are insufficient to ascertain the dynamical state for this cluster. However, it is likely that RX J1757, Cl 1604A and RX J0910+5419 are relaxed clusters.

While our evidence points to two clusters appearing unrelaxed, it is still possible that the offsets from the virialization relations could be the result of incomplete spectroscopy. This concern is validated by past observations of Cl 1324+3011, where \citet{lub04} measured a dispersion of $1016^{+126}_{-93}$ \kms\ using 47 galaxies within the entire LRIS field of view, well above the $\sigma_v$-T and $L_x$-$\sigma_v$ relations for virialized clusters. Since then, we have taken spectroscopic data on the Cl 1324 supercluster using 10 DEIMOS masks. Our most recent velocity dispersion measurement, shown in Table \ref{specanntab}, is $920\pm120$ \kms, using 45 galaxies within 1 $h_{70}^{-1}$ Mpc of the red galaxy density peak, offset from the $\sigma_v$-T relation by less than $1\sigma$. While we used fewer galaxies than \citet{lub04}, we measure the dispersion in a region consistent with the other clusters in our sample, and where the spectroscopic completeness is much improved. The new, more reliable set of cluster members had a significant effect on the velocity dispersion measurement and the scaling relations. The clusters with fewer high-quality spectra would be most prone to mismeasurement, such as those in the RX J0910 supercluster and Cl 1324+3013. 

\subsection{Implications for Cosmological Cluster Surveys}

The ORELSE survey has a multiwavelength dataset, including an unprecedented amount of spectroscopic data for clusters embedded in large-scale structure at high redshifts. These data have allowed us to implement a broader variety of diagnostics of the clusters' dynamical states than possible for most clusters at comparable redshifts. While many studies involving scaling relations attempt to identify unrelaxed clusters, dynamical states are often determined using only X-ray morphology of clusters \citep[e.g.,][]{pratt09,vik09b,maug11}. For our sample, we find that X-ray morphology alone is insufficient to identify the unrelaxed clusters. The clusters that we determined to be unrelaxed, Cl 1604B and RX J0910+5419, do not have the most asymmetric X-ray contours. While differing measures of asymmetry are used throughout the literature \citep[e.g.,][]{BT96,krav06,vik09b,hud10,maug11}, some clusters in our sample which appear reasonably relaxed (e.g., Cl 1324+3059 and Cl 1604A) would be excluded in most cases on X-ray morphology cuts, while our confirmed unrelaxed clusters might not be excluded. Highly exclusive cuts would be necessary to remove all unrelaxed clusters from our sample.

While current morphology cut techniques may be sufficient for some studies, the problem of efficiently determining dynamical states of clusters is important for cosmology. For example, measurements of $\sigma_8$ using galaxy cluster counts or the cluster mass function are impacted by improper corrections for unrelaxed clusters \citep[e.g.,][]{voit05}. In a recent study, \citet{vik09b} determined cluster masses using proxy properties. For unrelaxed clusters, the masses estimated using their M-$T_x$ relation were shifted upward by a constant factor relative to relaxed clusters. This correction is justified because unrelaxed clusters, defined using X-ray morphology, have been found through simulations to have masses 17$\pm$5\% higher than those of relaxed clusters for a given $T_x$ \citep{krav06}. Other studies have found similar results using different selection methods \citep[e.g.,][]{AS12}. With better selection of unrelaxed clusters, this correction, and the subsequent measurement of cosmological parameters, could be improved upon. The problem of unreliable mass measurements for unrelaxed clusters can also be bypassed by using the alternative mass proxy $Y_x$, which is the product of $T_x$ and $M_g$, the mass of the ICM gas, which is designed to be similar to and more easily measured than the Sunyaev-Zel'dovich (SZ) flux \citep{krav06}. \citet{krav06} find that the scatter in the $M_{500}$-$Y_x$ relation is approximately 6\%, compared to a $\sim 20\%$ scatter in the $M_{500}$-$T$ relation, without significant differences in mass measurements for relaxed and unrelaxed clusters. However, when only $T_x$, or even more problematic only $L_x$, is available, a better selection method is necessary to improve mass measurements for clusters. While the effect on measurements of $\sigma_8$ may not be dramatic, systematics dominate the error budget of measurements from cluster surveys \citep[e.g.,][]{vik09}. In the era of precision cosmology, reducing systematic errors is a worthwhile and important goal. 

While our sample has demonstrated the deficiencies of selecting unrelaxed clusters based on X-ray morphology alone, it is too small to determine ways to significantly improve this method or to do more than provide a selection of possible alternatives, such as the offset between the X-ray centroid and the red galaxy density peak that appears to be indicative of an unrelaxed cluster (see section \ref{sec:anal}). However, the full ORELSE survey could supply the sample necessary to improve the selection of unrelaxed clusters. The ORELSE survey has chosen 20 clusters around which to search for large-scale structure. Only six of these have been studied in detail\footnote{In addition to the five fields covered in this paper, the Cl0023 supergroup has also been studied in detail, although only groups lacking detected diffuse X-ray emission have been identified in the structure.}. With detailed studies of the remaining 14 fields, and more data on some of the fields in this paper, many more clusters will be added to our sample. In addition, clusters from other sources with high-redshift multiwavelength datasets can be combined, such as from the MAssive Cluster Survey \citep{eb01,eb10}, the Red-Sequence Cluster Surveys \citep{GY05,gil11}, and samples selected with the {\it Spitzer} Infrared Array Camera (IRAC) \citep[e.g.,][]{eis08,krick08}. These larger samples will provide even better potential for analysis of techniques of identifying relaxed clusters. 

\section{Conclusions}
\label{sec:con}

We presented the results of a search for diffuse X-ray emission from clusters and groups in five large-scale structures at high redshifts using deep {\it Chandra} imaging. We detected emission from a total of nine clusters and were able to measure gas temperatures for eight of these. While we detected emission from the two isolated clusters, RX J1757 and RX J1821, we only detected emission from two clusters in the Cl 1604 supercluster and three clusters in the Cl 1324 supercluster. We also detected emission from RX J0910+5419 and RX J0910+5422. While there is evidence from infrared imaging of additional structure in the RX J0910 field \citep{tan08}, we do not detect emission from any other groups or clusters. 

Except for Cl 1604B and RX J0910+5419, we found that all clusters with detected diffuse emission were consistent with velocity dispersion, gas temperature, and bolometric X-ray luminosity scaling relations for low-$z$ virialized clusters at the $2\sigma$ level, although two clusters in Cl 1324 and RX J1821 may have inflated velocity dispersions, possibly due to filamentary structure or recent mergers. Cl 1604B and RX J0910+5419 were offset from the scaling relations, although by modest amounts, which implies that these systems are still in the process of gravitationally heating. For Cl 1604B, this result is supported by analysis of the velocity dispersions of red and blue populations. We find minimal differences between the velocity dispersions of red and blue galaxies, less than for any other cluster, suggestive of a younger cluster and potential substructure \citep{ZF93,gal08}. We also found a significant difference between the velocity centers of these two populations, suggesting the presence of substructure \citep{ZF93}. Due to an insufficient number of spectroscopically confirmed members, we were unable to carry out this analysis of red and blue galaxies for RX J0910+5419, as well as for three other clusters in our sample. The BCG in RX J0910+5419 has a large peculiar velocity, and both this cluster and Cl 1604B have large offsets between their red galaxy density peaks and the X-ray emission, which provides further evidence against virialization.


Several studies have found evidence for evolution in the virial relations. \citet{ett04} and \citet{hicks08} both find an increase in the exponent of the $L_x$-$T$ relation at higher redshifts, so that clusters at the same temperature have lower luminosities. \citet{rei11} have compiled data from a number of studies to determine the redshift evolution of scaling relations, finding a deviation from self-similarity in the $L_x$-T relation. While our own sample is too small to fit for redshift evolution, we find our data to be consistent both with these previous studies and with self-similarity. However, as shown in Figure \ref{LxTevolfig}, all of the clusters that appear to be reasonably relaxed are below the line representing self-similar evolution, suggesting it is not characteristic of our data. A larger sample, a larger redshift range, or both may be necessary to fit the redshift evolution of this relation or to better evaluate the relations of others. 

A larger sample could also assist in better understanding tests of dynamical state. While many studies use cluster X-ray morphology as a measure of dynamical state \citep[e.g.,][]{pratt09,maug11}, we find it is a poor predictor of unrelaxed clusters for our sample. While morphology tests may still be used as a means of excluding unrelaxed clusters from surveys, the biases of these techniques, or any used in their place, need to be better understood with a larger sample. Properly identifying unrelaxed clusters could have an impact on measurements of cosmological parameters using galaxy cluster counts surveys, which are currently dominated by systematic errors. 


This work is supported by the Chandra General Observing Program under award numbers GO6-7114X, GO7-8126X, GO8-9123A, and GO9-0139A. In addition, we acknowledge support by the National Science Foundation under grant AST-0907858. Spectrographic data presented herein were obtained at the W.M. Keck Observatory, which is operated as a scientific partnership among the California Institute of Technology, the University of California, and the National Aeronautics and Space Administration. The Observatory was made possible by the generous financial support of the W.M. Keck Foundation. As always, we thank the indigenous Hawaiian community for allowing us to be guests on their sacred mountain. We are most fortunate to be able to conduct observations from this site.

\appendix
\section{X-ray Count Profiles}
\label{sec:SB}

\begin{figure*}
\epsscale{1.1}
\plotone{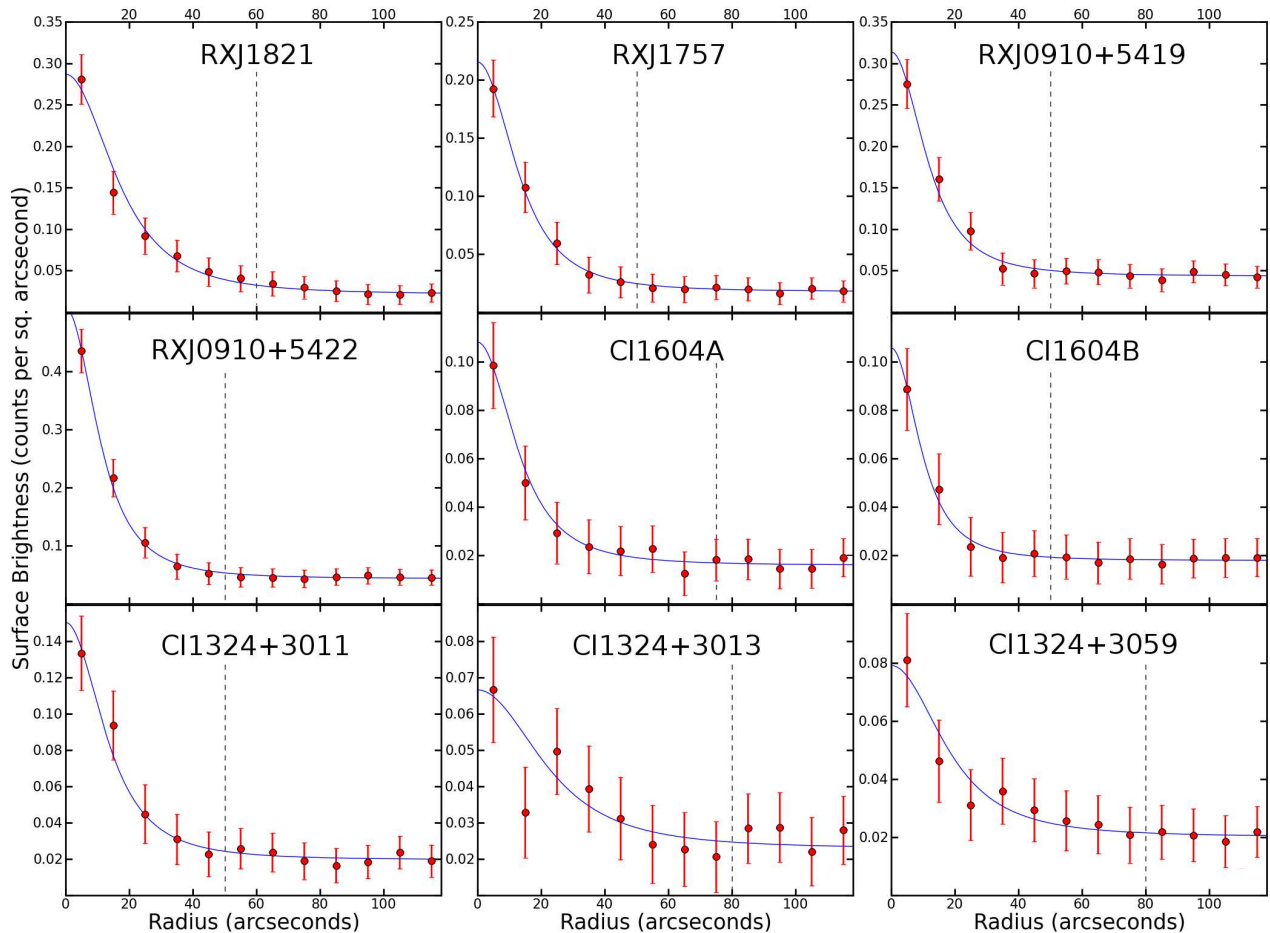}
\caption{\footnotesize{
Soft band (0.5-2.0 keV) X-ray surface brightness profiles for each cluster are displayed without subtraction of the background. Each point represents {\it Chandra} X-ray counts within an annulus divided by that annulus' area in square arcseconds. Vertical dashed lines indicate the radius in which spectra were extracted. Beta model fits to the data are also shown with solid lines. 
}}
\label{SBfig}
\end{figure*}

As mentioned in Section \ref{sec:DE}, X-ray surface brightness\footnote{Here, surface brightness is defined as the X-ray photon counts per square arcsecond.} plots were used for each cluster to determine the area in which to measure X-ray counts and the temperatures. Surface brightnesses were measured in concentric annuli centered on the center of X-ray emission for each cluster. The surface brightness profiles are shown in Figure \ref{SBfig}. Each point represents the X-ray counts in the 0.5-2.0 keV band in an annulus divided by the area of that annulus in square arcseconds. The point is shown halfway between the inner and outer radii of the annulus. The lines represent fits to the data, which will be explained later in the section. We can see that most of the surface brightness profiles asymptoticly approach some background level. For each cluster, a radius was chosen at which the surface brightness had approximately reached the background. X-ray counts and the temperatures of the clusters were measured inside these extraction radii. For Cl 1324+3013, the surface brightness falls with increasing radius, but then rises again, which is due to the proximity of a foreground cluster (see Section \ref{sec:DE}). The extraction radius for Cl 1324+3013 was chosen to be 80 arcseconds, where the surface brightness levels off but before it rises again. The extraction radii for all clusters are listed in Table \ref{specanntab}. 

For the clusters in our sample, the extraction radii correspond to differing physical distances. For this reason, luminosities were extrapolated to $r_{500}$ for comparison. In order to extrapolate, fits were performed to the one-dimensional surface brightness plots shown in Figure \ref{SBfig}. For our azimuthally averaged surface brightness model, we used a beta model plus a constant background: \begin{equation}
SB\left(r\right) = A \left(1+r^2/r_c^2\right)^{1/2-3\beta} + SB_{bkg}
\end{equation}
We set $\beta = 2/3$, which is a typical value for clusters \citep[see e.g.,][]{AE99,ett04,maug06,hicks08}. We also required the net photon counts, $NC$, from our model within the extraction radius, $r_e$, to be equal to that measured using our {\it Chandra} data. This requirement translates to \begin{equation}
NC_{r_e} = 2\pi A r_c^2\left(1-1/\sqrt{1+r_e^2/r_c^2}\right)
\end{equation}
which creates a relation between the core radius, $r_c$, and the normalization, $A$, and reduces our model to two parameters: $r_c$ and $SB_{bkg}$. With this constraint, we fit our model using Sherpa and $\chi^2$ statistics. Fitted core radii ranged from 100-210 $h_{70}^{-1}$\ kpc, which are typical values for clusters \citep[see e.g.,][]{ett04,maug06,hicks08}. The fits to the data are shown in Figure \ref{SBfig}. Note that for Cl 1324+3013, the data points beyond 80 arcseconds were not used for the fit due to the contamination from the foreground cluster. 

\end{document}